\documentclass[symmetry,review,accept,pdftex,moreauthors]{Definitions/mdpi}

\graphicspath{{figures/}} 

\firstpage{1}
\pubvolume{15}
\issuenum{3}
\articlenumber{727}
\pubyear{2023}
\copyrightyear{2023}
\externaleditor{{Academic Editors: Vasilis K. Oikonomou and Silvio Pardi}}
\datereceived{31 December 2022}
\daterevised{17 February 2023}
\dateaccepted{6 March 2023}
\datepublished{15 March 2023}
\hreflink{https://doi.org/10.3390/\linebreak sym15030727} 

\Title{Heavy-Flavour Jets in High-Energy Nuclear Collisions}
\TitleCitation{Heavy-Flavour Jets in High-Energy Nuclear Collisions}

\Author{{Sa Wang} $^{1,2}$\orcidA{}, Wei Dai $^{3}$, Enke Wang $^{1,2}$, {Xin-Nian Wang} $^{4}$ and {Ben-Wei Zhang} $^{2}$*}

\AuthorNames{Sa Wang, Wei Dai, Enke Wang, Xin-Nian Wang, Ben-Wei Zhang}

\AuthorCitation{{Wang, S.; Dai, W.; Wang, E.; Wang, X.-N.; Zhang, B.-W.}}

\address{%
$^{1}$ \quad Guangdong Provincial Key Laboratory of Nuclear Science, Institute of Quantum Matter,\linebreak South China Normal University, Guangzhou 510006, China\\
$^{2}$ \quad {Key Laboratory of Quark \& Lepton Physics (MOE)} and {Institute of Particle Physics},\linebreak Central China Normal University, Wuhan 430079, China\\

$^{3}$ \quad School of Mathematics and Physics, China University of Geosciences, Wuhan 430074, China\\
$^{4}$ \quad {Nuclear Science Division MS 70R0319, Lawrence Berkeley National Laboratory}, Berkeley,  CA 94720, USA\\}

\corres{Correspondence: bwzhang@mail.ccnu.edu.cn}

\abstract{Reconstructed jets initiated from heavy quarks provide a powerful tool to probe the properties of the quark-gluon plasma (QGP) and to explore the mass hierarchy of jet quenching. In this article, we review the recent theoretical progresses on heavy-flavour jets in high-energy nuclear collisions at the RHIC and LHC. We focus on the yields and substructures of charm and bottom quark jets with jet quenching effect, such as the nuclear modification factors, transverse momentum imbalance, angular correlation, radial profiles, fragmentation functions, the ``dead-cone'' effect, etc.}


\begin{document}


\section{Introduction}
\label{sec:intro}

High-energy nuclear collisions at the Relativistic Heavy Ion Collider~(RHIC) and Large Hadron Collider~(LHC) have opened up new avenues for the search for strongly interacting nuclear matter, the quark--gluon plasma~(QGP)~\cite{Wang:1992qdg, Gyulassy:2003mc, Mehtar-Tani:2013pia, Qin:2015srf}. Investigating the formation of the QGP deepens our understanding of quantum chromodynamics (QCD) under extreme conditions at high temperature and density~\cite{Cunqueiro:2021wls, Cao:2020wlm} and the evolution of the Universe at the first microsecond~\cite{Collins:1974ky}. The jet-quenching phenomena, the energy attenuation of fast partons due to their strong interactions with the QCD medium, provide an army of powerful tools to study the properties of the QGP, such as the yield suppression of high-$p_T$ hadron/jet, the $p_T$ asymmetry of dijets, $\gamma/Z^0$+ jets as well as jet substructures~\cite{Wang:1998bha,Wang:2001cs,Wang:2002ri,Majumder:2004pt,Zhang:2007ja,Vitev:2009rd,He:2020iow,Wang:1996yh,Neufeld:2010fj,Dai:2012am,Wang:2013cia,Li:2010ts,Vitev:2008rz,Caucal:2021cfb}.

In elementary proton--proton reactions, the productions of charm and bottom quarks are perturbatively calculable, since their large masses~($M_c\sim 1.5$~GeV, $M_b\sim 4.8$~GeV) act as a natural cut-off above the $\Lambda_{QCD}$~\cite{Andronic:2015wma}. Heavy quarks are produced in the initial hard scattering at a very early stage due to their large masses, therefore witnessing the whole QGP evolution. Meanwhile, while their thermal production is almost negligible with the initial conditions so far accessible in heavy-ion programs at the RHIC and LHC~\cite{Zhang:2007yoa}, the productions of charm and bottom hadron/jets make a very promising hard probe to the transport properties of hot and dense quark matter. During the past decade, the experimental measurements including the nuclear modification factor $R_{AA}$~\cite{Adamczyk:2014uip,Adam:2015sza,Sirunyan:2017xss,PHENIX:2011img,STAR:2013eve,ALICE:2014wnc,CMS:2011all} and the collective flow (the direct flow $v_1$~\cite{STAR:2019clv,ALICE:2019sgg} and elliptical flow $v_2$~\cite{Abelev:2014ipa,Adamczyk:2017xur,Acharya:2017qps,Sirunyan:2017plt}) of heavy-flavour hadrons both at the RHIC and LHC have attracted much attention from the community of high-energy nuclear physics.

A lot of theoretical studies have been performed to confront the experimental data obtained in the high-energy heavy-ion collisions, which greatly improve our understanding of the in-medium evolution~\cite{Zhang:2003wk,Djordjevic:2003zk,Zhang:2004qm,vanHees:2007me,CaronHuot:2008uh,Djordjevic:2015hra,He:2014cla,Kang:2016ofv,Svetitsky:1987gq,Moore:2004tg,
Cao:2013ita,Alberico:2013bza,Xu:2015bbz,Cao:2016gvr,Das:2016cwd,Ke:2018tsh,Li:2020kax,Liu:2009nb,Yan:2006ve,JETSCAPE:2022hcb,Liu:2021dpm,Ding:2021ajz} and hadronization mechanisms~\cite{Plumari:2017ntm,Cao:2019iqs,He:2019vgs} of heavy quarks (for detailed reviews see~\cite{Rapp:2009my,Prino:2016cni,Rapp:2018qla,Dong:2019unq,Dong:2019byy,Zhao:2020jqu,Apolinario:2022vzg,He:2022ywp,Tang:2020ame}). Specifically, the current models treat the elastic and inelastic interactions between heavy quarks and the QGP medium with multiple methods, consisting mainly the perturbative or non-perturbative analytic calculations (SCET~\cite{Kang:2016ofv,Li:2018xuv}, \textls[-25]{CUJET~\cite{Xu:2015bbz,Xu:2014tda}, DREENA~\cite{Djordjevic:2009cr,Djordjevic:2008iz,Djordjevic:2013xoa,Zigic:2021rku,Zigic:2018ovr,Zigic:2018smz}, WHDG~\cite{Wicks:2007am,Wicks:2005gt}, AdS/CFT (HG)~\cite{Horowitz:2007su,Horowitz:2011wm}), and the Monte Carlo transport approaches based on the Boltzmann (\mbox{BAMPS~\cite{Uphoff:2010sh,Uphoff:2011ad,Uphoff:2013rka,Uphoff:2014hza}}, \textls[-5]{MC$@_s$HQ~\cite{Gossiaux:2008jv,Gossiaux:2009mk,Nahrgang:2013saa}, \linebreak (Q)LBT~\cite{Cao:2016gvr,Liu:2021dpm}, LIDO~\cite{Ke:2018tsh,Ke:2020clc}, Catania-pQCD/QPM~\cite{Plumari:2012ep,Das:2015ana,Plumari:2011mk,Scardina:2017ipo}), the Langevin} (POWL \linebreak ANG~\cite{Alberico:2013bza,Alberico:2011zy,Beraudo:2014boa},} Duke~\cite{Cao:2013ita,Cao:2011et}, UrQMD~\cite{Lang:2012nqy,Lang:2013cca,Lang:2013wya}, TAMU~\cite{He:2012df,He:2011yi,He:2014cla}, \mbox{SHELL~\cite{Wang:2019xey,Dai:2018mhw,Wang:2021jgm}}) and the Kadanoff--Baym (PHSD~\cite{Cassing:2008sv,Cassing:2009vt,Bratkovskaya:2011wp}) equations. These phenomenological studies reveal a fact that the elastic scattering of heavy quarks in the hot/dense nuclear matter is important, especially at the lower $p_T$ region ($p_T^Q<5m_Q$), different from our experience of treating light quarks or gluons. One of the central issues of investigating the heavy-flavour production in the heavy-ion program is extracting the diffusion coefficient $D_s$, which is directly related to the transport properties of the hot QCD matter. Additionally, different from the fragmentation hadronization of heavy quarks in a vacuum, within the hot and dense nuclear matter, the heavy-flavour hadrons can be produced by a combination of heavy quarks and thermal partons. Such a coalescence hadronization mechanism plays an important role in the collective flow~\cite{Abelev:2014ipa,Adamczyk:2017xur} and baryon-to-meson ratio~\cite{STAR:2019ank,Vermunt:2019ecg} of charmed hadron in nucleus--nucleus collisions at the RHIC and LHC.

In recent years, the experimental measurements on heavy-flavour jet (a reconstructed jet containing a heavy quark or a heavy-flavour hadron) have made great strides in \linebreak p+p~\cite{CMS:2016wma,ALICE:2019cbr,CMS:2012pgw,ATLAS:2011chi,ATLAS:2021agf,STAR:2009kkp,ATLAS:2016anw,CMS:2020geg}, p+A~\cite{CMS:2016wma,Khachatryan:2015sva,ALICE:2021wct} and A+A collisions~\cite{Chatrchyan:2013exa,Sirunyan:2018jju,Sirunyan:2019dow,ATLAS:2022fgb,Roy:2022yrw,CMS:2021puf,ALICE:2018lyv,CMS:2022btc}. A wealth of information carried by heavy-flavour jets not only offers a new topic of jet physics and the application of the perturbative QCD, but also their medium modifications in heavy-ion collisions are also of great significance to reveal the in-medium energy loss mechanism of heavy quarks, to address the mass effect of jet quenching, and to extract the transport properties of the~QGP.

\section{Recent Advances of Heavy-Flavour Phenomenology in Heavy-Ion Collisions}
\label{sec:overview}

Generally speaking, as we discussed in the last section, the reason for treating the heavy flavours as powerful hard probes to the transport properties of the QGP consists of at least three aspects. Firstly, the large mass ($M_Q \gg \Lambda_{QCD}$) makes it available to compute the differential cross-section of heavy quarks in the binary nucleon--nucleon collisions based on the perturbative QCD (pQCD) scheme within the next-to-next-to-leading order (NNLO) precision~\cite{Cacciari:2005rk}. Secondly, due to the large mass ($M_Q \gg T_{\rm med}$), the total yield of heavy quarks in nucleus--nucleus collisions only depends on their initial production at hard scattering. Since the momentum transfer of the in-medium collisions $q^2\sim g^2T^2$\linebreak ($T\sim 0.4$--$0.5$ GeV) is much smaller than the creation energy of heavy quark pairs at the current collision energy, both at the RHIC and LHC, the subsequent contribution from the thermal creation during the QGP evolution is negligible~\cite{Zhang:2007yoa}. Apart from this, according to the Heisenberg uncertainty principle, the formation time of heavy quarks\linebreak ($\tau_0\sim \frac{1}{2m_Q}<0.1$~fm/c) is shorter than the formation time of the quark--gluon plasma ($\tau_f\sim 0.6$ fm/c). Therefore heavy quarks witness the entire evolution of the hot/dense nuclear matter until the freeze-out. In this section, we will briefly introduce the recent theoretical advances that help us understand the heavy-flavour production in heavy-ion collisions, including mainly the following several aspects, the initial production, the transport approaches, the hadronization mechanisms, and the extraction of diffusion~coefficient.

\subsection{Production of Heavy Quarks in p+p Collisions}
\label{sec:ppbaseline}

The production of heavy quarks in proton--proton collisions establishes a baseline to investigate the nuclear modification in high-energy nuclear collisions both at the RHIC and LHC. The yield of heavy flavours in nucleus--nucleus collisions generally is viewed as the sum of that in $N_{\rm coll}$ binary nucleon--nucleon collisions while taking into account the initial cold nuclear matter effect (usually considered by using the nuclear-modified parton distribution function~\cite{Eskola:2009uj,Eskola:2016oht,NNPDF:2014otw}). In the fixed-flavour-number scheme (FFNS)~\cite{Andronic:2015wma}, the cross-section of heavy quarks in p+p collisions can be expressed based on the factorization~theorem,
\vspace{-10pt}
\begin{adjustwidth}{-\extralength}{0cm}
\centering 
\begin{eqnarray}
d{\sigma}_Q[s,p_T,y,m_Q]\simeq\sum_{i,j}\int_0^1 dx_i\int_0^1 dx_j f_i^A(x_i,\mu_F)f_j^A(x_j,\mu_F)d\tilde{{\sigma}}_{ij\rightarrow Q+X}[x_i,x_j,s,p_T,y,m_Q,\mu_F,\mu_R]
\label{eq:sigma}
\end{eqnarray}
\end{adjustwidth}
where $s$ is the square of the centre-of-mass energy of the incoming proton, $p_T$ is the transverse momentum of the produced heavy quark, and $y$ is the rapidity. $f_i^A$ ($f_j^B$) is the parton distribution function (PDF) quantifying the probability to find a parton with flavour {$i$($j$)} 
and carrying momentum fraction $x_{i(j)}$ in the colliding proton $A(B)$, which relies on the factorization scale $\mu_F$. {$\tilde{{\sigma}}_{ij\rightarrow Q+X}$} represents the cross-section of the partonic hard process $i+j\rightarrow Q+X$ that can be calculated relying on the pQCD. 
The partonic cross-section $\tilde{{\sigma}}_{ij\rightarrow Q+X}$ also relies on the strong coupling constant $\alpha_s$ determined at the renormalization scale $\mu_R$. Note that Equation~(\ref{eq:sigma}) sums all partonic hard processes $i+j\rightarrow Q+X$, where $i,j$ are the active flavours including ($u,\bar{u},d,\bar{d},s,\bar{s},g$) but not heavy quarks. Only at the factorization scale $\mu_F>m_c$, can charm be viewed as an active flavour, often used for beauty production. The differential cross-section $d{\sigma}_Q$ can be convolved with a scale-independent fragmentation function $D_Q^H(z)$, such as the Peterson~\cite{Peterson:1982ak} or Lund~\cite{Andersson:1983ia} forms, to obtain the cross-section of the heavy-flavour hadron,
\begin{eqnarray}
d{\sigma}_H=d{\sigma}_Q\otimes D_Q^H(z)
\label{eq:sigmaH1}
\end{eqnarray}
where $H$ denotes the heavy-flavour hadron and $z$ the momentum fraction carried by $H$.

Since the FFNS is usually applicable at the low $p_T$ region ($0<p_T<5m_Q$), for the higher kinematic region ($p_T\gg m_Q$), the logarithmic terms ($\frac{\alpha_s}{2\pi}ln(p_T^2/m_Q^2)$) in the perturbative expansion of the cross-section become large, and should be resummed to all orders. To implement such a resummation, one has to absorb the large logarithmic terms into the parton distribution function and fragmentation function. This treatment requires that heavy quarks are active flavours when the factorization scale is $\mu_F>m_Q$. In other words, such a scheme has a variable number of active flavours when $\mu_F$ crosses the heavy quark mass, hence named the variable-flavour-number scheme (VFNS). In particular, when the heavy quark mass can be neglected in the evaluation of the short-distance cross-section, the VFNS scheme is called the zero-mass VFNS (ZM-VFNS). In the ZM-VFNS, the differential cross-section of a heavy-flavour hadron based on the factorization theorem can be expressed~as:
\begin{eqnarray}
d{\sigma}_{H+X} \simeq \sum_{i,j}\int_0^1 dx_i\int_0^1 dx_j f_i^A(x_i,\mu_F)f_j^A(x_j,\mu_F)d\tilde{{\sigma}}_{ij\rightarrow k+X}D_k^H(z,{\mu}_F^{\prime})
\label{eq:sigmaH2}
\end{eqnarray}
where $D_k^H(z,{\mu}_F^{\prime})$ is given by a convolution of a perturbative-fragmentation function (PFF) $D_k^Q(z,{\mu}_F^{\prime})$ describing a parton k fragmentation into heavy quark Q, with a scale-independent one $D_Q^H(z)$ for the hadronization of a heavy quark. Note that in Equation~(\ref{eq:sigmaH2}) the sum covers all possible partonic-hard processes ($i+j\rightarrow k+X$) where $i,j,k $ can be light quarks, gluons, and heavy quarks~\cite{Aversa:1988vb}. Since heavy quark mass is neglected in the computation of the cross-section, the ZM-VFNS is expected to be reliable only at very high~$p_T$.

To find a unified theoretical framework that combines the advantages of the FFNS at low $p_T$ and the ZM-VFNS at high $p_T$, in recent years the interpolation schemes have been established, such as the general-mass VFNS (GM-VFNS)~\cite{Kniehl:2004fy,Kniehl:2005mk} and the fixed-order plus next-to-leading logarithms (FONLL)~\cite{Cacciari:2005rk,Cacciari:2012ny}. For instance, by using an interpolating function $G(m_Q,p_T)=p_T^2/(p_T^2+c^2m_Q^2)$ where $c$ is set to $c=5$, the FONLL scheme can well describe the heavy-flavour production in the entire kinematic region. For more details of the interpolation schemes see~\cite{Andronic:2015wma} and the references therein.

Compared to the analytic calculation schemes discussed above, the general-purpose Monte Carlo event generator, such as PYTHIA~\cite{Sjostrand:2006za}, HERWIG~\cite{,Corcella:2000bw}, POWHEG~\cite{Frixione:2007vw} and SHERPA~\cite{Gleisberg:2008ta}, can provide a more complete description of all the final-state particles at the parton or hadron level. Especially for the studies of jet physics, the Monte Carlo event generators can give more precise descriptions of the observations relating to the jet substructure than analytic calculations.

\subsection{Transport of Heavy Quarks in the QGP}
\label{sec:transport}

Transport approaches are wildly used in the current theoretical studies of heavy-flavour production in high-energy nuclear collisions. At the lower $p_T$ region, the elastic scattering of heavy quarks with the thermal parton (light quark or gluon) has been proven to be the dominant mechanism of energy loss. Generally,  the kinetic theory based on the Boltzmann transport equation is a popular treatment for in-medium heavy quark evolution. The Boltzmann equation for the distribution function of heavy quarks can be written in a compact form,
\begin{eqnarray}
p_{\mu}\partial_{\mu}f_Q(x,p)=\mathit{C}[f_q,f_{\bar{q}},f_g,f_Q](x,p)
\label{eq:Boltzmann}
\end{eqnarray}
where $f_Q(x,p)$ is the phase-space distribution of heavy quarks. In the QGP, the phase-space distributions of light quark $f_q$ and gluon $f_g$ can be solved by the Boltzmann equation~\mbox{\cite{Ferini:2008he,Ruggieri:2013ova}}. Subsequently, the relativistic Boltzmann-like collision integral $\mathit{C}[f_Q](x,p)$ has a simplified form~\cite{Svetitsky:1987gq,Beraudo:2014iva},
\begin{eqnarray}
\mathit{C}[f_Q]=\int d^3q[\omega(\mathbf{p}+\mathbf{q},\mathbf{q})f_Q(\mathbf{x},\mathbf{p}+\mathbf{q},t)-\omega(\mathbf{p},\mathbf{q})f_Q(\mathbf{x},\mathbf{p},t)]
\label{eq:integ}
\end{eqnarray}
{where} $\omega(\mathbf{p}+\mathbf{q},\mathbf{q})$ represents the transition rate of a heavy quark from the momentum $\mathbf{p}+\mathbf{q}$ to $\mathbf{p}$ by collisions with quasiparticles. This rate is usually determined by the matrix elements of the $2\rightarrow 2$ QCD scattering. With the assumption that the momentum transfer $|\mathbf{q}|$ is small compared to the momentum of a heavy quark, we can expand $\omega(\mathbf{p}+\mathbf{q},\mathbf{q})f_Q(\mathbf{x},\mathbf{p}+\mathbf{q},t)$ around $\mathbf{q}$ by utilizing the Taylor formula to obtain the Fokker--Planck equation,
\begin{eqnarray}
\frac{\partial f_Q}{\partial t}=\frac{\partial}{\partial p_i}\left [ A_i(\mathbf{p})f_Q+\frac{\partial}{\partial p_j}[B_{ij}(\mathbf{p})f_Q] \right ]
\label{eq:FP}
\end{eqnarray}
where two coefficients $A_i(\mathbf{p})=\int d^3q\omega(\mathbf{p},\mathbf{q})q_i$ and $B_{ij}(\mathbf{p})=\int d^3q\omega(\mathbf{p},\mathbf{q})q_iq_j$ are directly related to the drag coefficient ($\eta_D$) and the momentum diffusion coefficient ($\kappa$), which control rate of the energy loss and the momentum broadening of heavy quarks in the hot medium, respectively. Indeed, the Fokker--Planck equation is equivalent to another more well-known equation, the Langevin equation,
\begin{eqnarray}
\frac{d\vec{x}}{dt}&=&\frac{\vec{p}}{E}\\
\frac{d\vec{p}}{dt}&=&-\eta_D(p) \vec{p}+\vec{\xi}(t)
\label{eq:Fokker}
\end{eqnarray}
where the stochastic term $\vec{\xi}(t)$ describes the random kicks suffered in heavy quarks from the medium constituents, which obeys a Gaussian distribution with a mean value $0$ and variance $\kappa$. The drag coefficient $\eta_D$ and the diffusion coefficient $\kappa$ are related by the fluctuation--dissipation theorem (FDT) $\kappa=2\eta_DET$. Note that at higher kinematic regions ($p_T^Q>5m_Q$), the medium-induced gluon radiation plays an increasingly important role in the energy loss of heavy quarks. The radiative energy loss of heavy quarks is treated with various formalisms and at different approximations~\cite{Zhang:2003wk,Arnold:2002ja,Abir:2011jb,Abir:2012pu,Zapp:2008gi,Armesto:2003jh}, which usually provide the radiated gluon spectra as a function of momentum fraction $x$ and transverse momentum $k_{\perp}$. In the Langevin equation, the radiative energy loss of heavy quarks can be coupled with the collisional one by adding a recoil term $-\vec{p}_g$ caused by the radiated gluon~\cite{Cao:2013ita}. The four-momentum of the radiated gluon can be easily sampled based on the radiation spectra $d{\rm N}_g/d{\rm x}d{\rm k}_{\perp}^2$.

In many of the recently developed theoretical frameworks modelling the production of heavy flavour in heavy-ion collisions, the Boltzmann and Langevin equations are the two most popular choices, especially for Monte Carlo simulations. Concerning the performance of these two approaches, detailed comparisons have been discussed in~\cite{Das:2013kea, Li:2019wri}. In general, the implementation of the Boltzmann equation implies that the medium consists of well-defined quasiparticles, while the Fokker--Planck (Langevin) equation is realized in a more general way without the quasiparticle assumption. However, the advantage of the Boltzmann equation is that it can naturally describe the heavy quark evolution even under off-equilibrium conditions, which may be the case of the early pre-equilibrium stage in heavy-ion collisions~\cite{Das:2017dsh}.

\subsection{Hadronization: Fragmentation and Coalescence}
\label{sec:hadronization}
Studying the yield suppression and collective flow of heavy-flavour hadrons also deepens our understanding of heavy quark hadronization in nucleus--nucleus collisions, which shows different mechanisms with that in a vacuum. As discussed in Section~\ref{sec:ppbaseline}, fragmentation functions describe the non-perturbative hadronization process of heavy quarks into heavy-flavour hadron in a vacuum. The most commonly used fragmentation function is the Peterson form~\cite{Peterson:1982ak},
\begin{eqnarray}
D_{H/Q}(z)=\frac{N}{z[1-\frac{1}{z}-\frac{\epsilon_Q}{1-z}]}
\label{eq:frag}
\end{eqnarray}
where $z$ denotes the momentum fraction carried by the heavy hadron from the heavy quark in the fragmentation process ($0<z<1$), which implies that the heavy hadron must have smaller energy than the heavy quark. The only tunable parameter in Equation~(\ref{eq:frag}) is $\epsilon_Q$ that can be determined by fitting to the measured spectra of the heavy-flavour hadrons. $N$ is the normalization factor to guarantee $\int_0^1 dz D_{H/Q}(z)=1$.

Measurements on the collective flow~\cite{Abelev:2014ipa, Adamczyk:2017xur} and baryon-to-meson ratio~\cite{STAR:2019ank, Vermunt:2019ecg} of charmed hadron A+A collisions suggest the existence of a new hadronization mechanism, coalescent of heavy quarks. 
The basic idea behind the coalescence mechanism is that a heavy quark can combine with a light anti-quark from the medium when they have enough small distance in the coordinate-momentum space. It means that the heavy-flavour meson has larger energy than the parent heavy quark, differing from the mechanism of fragmentation. The distribution function of the formed heavy-flavour meson usually can be obtained by a convolution with the following schematic form.
\begin{eqnarray}
f_M \sim g_M f_{Q(\bar{Q})} \otimes f_{\bar{q}(q)} \otimes \phi_M
\label{eq:coal}
\end{eqnarray}
where $g_M$ denotes the degeneracy of the heavy-flavour meson in spin and isospin, $f_{Q(\bar{Q})}$ and $f_{\bar{q}(q)}$ are the distribution functions of the heavy and light quarks in the coordinate-momentum space, respectively. $\phi_M$ represents the Wigner transform of the wave function of the heavy-flavour meson, commonly approximated by the ground state one of the simple harmonic oscillators~\cite{Cao:2016gvr}.

In the realistic implementation of the heavy quark hadronization in nuclear collisions, the first step is to determine the probability of coalescence by integrating the distribution function of Equation~(\ref{eq:coal}). If coalescence occurs, one can sample a light anti-quark based on the thermal equilibrium distribution, otherwise Equation~(\ref{eq:frag}) is used to fragment the heavy quark into a hadron. At least in the lower $p_T$ region, the experimental results favour the coalescence mechanism~\cite{Scardina:2017ipo}. The coalescence of heavy quarks seems to decrease the suppression factor and enhance the collective flow of heavy-flavour hadrons, especially at $p_T<6$ GeV. The recent studies~\cite{Cao:2019iqs,He:2019vgs} show that the coalescence mechanism is important in the description of the $\Lambda_c/D^0$ ratio measured by the STAR~\cite{STAR:2019ank} and ALICE~\cite{Vermunt:2019ecg} collaborations. Additionally, the hadronic scattering between the D meson and light-flavour hadrons ($D-\pi$, $D-\rho$) has also been studied in~\cite{Lin:2000jp}, but its influence on the D meson $R_{AA}$ was found to be very limited~\cite{Cao:2015hia}.

\subsection{Extraction of the Diffusion Coefficient of Heavy Quarks}
\label{sec:Extraction}

One of the most important goals of the heavy-ion collision experiment is to investigate the transport properties of the QCD matter under extremely hot and dense conditions. As discussed above, due to the large mass ($m_Q \gg T_{\rm med}$), heavy quarks are believed to be powerful tools for exploring the transport properties of the QGP. Phenomenological studies of heavy-flavour production in high-energy nuclear collisions provide a unique opportunity to extract the transport coefficient of the QGP, such as the momentum diffusion coefficient $\kappa$ of heavy quarks, whose longitudinal and transverse components can be convenient to define as,
\begin{eqnarray}
\kappa_{||}\equiv-\frac{d\left\langle (\Delta p_{||})^2 \right\rangle}{dt}\\
\kappa_{\perp}\equiv\frac{1}{2}\frac{d\left\langle (\Delta p_{\perp})^2 \right\rangle}{dt}
\label{eq:kappa}
\end{eqnarray}
where $\Delta p_{||}$ and $\Delta p_{\perp}$ momentum changes parallel and perpendicular to the heavy quark formulation. 
By definition, $\kappa_{\perp}$ can be directly related to the jet transport coefficient $\hat{q}$ which quantifies all the transverse momentum broadening of hard partons as traversing the QGP medium. Assuming that the $\kappa$ is isotropic, namely, $\kappa_{\perp}=\kappa_{||}=\kappa$, one can obtain a simplified relation $\hat{q}=2\kappa$. This relation has been employed in the modified Langevin equation to balance the two parts of the contribution from the collisional and radiative energy loss of heavy quarks~\cite{Cao:2013ita,Li:2019lex}. Here we only overview the recent advances of the $\kappa$ extraction by different model calculations. A more detailed and profound discussion about this topic can be found in~\cite{Rapp:2018qla}.

\begin{figure}[H]
\includegraphics[width=4.5in,height=3.7in,angle=0]{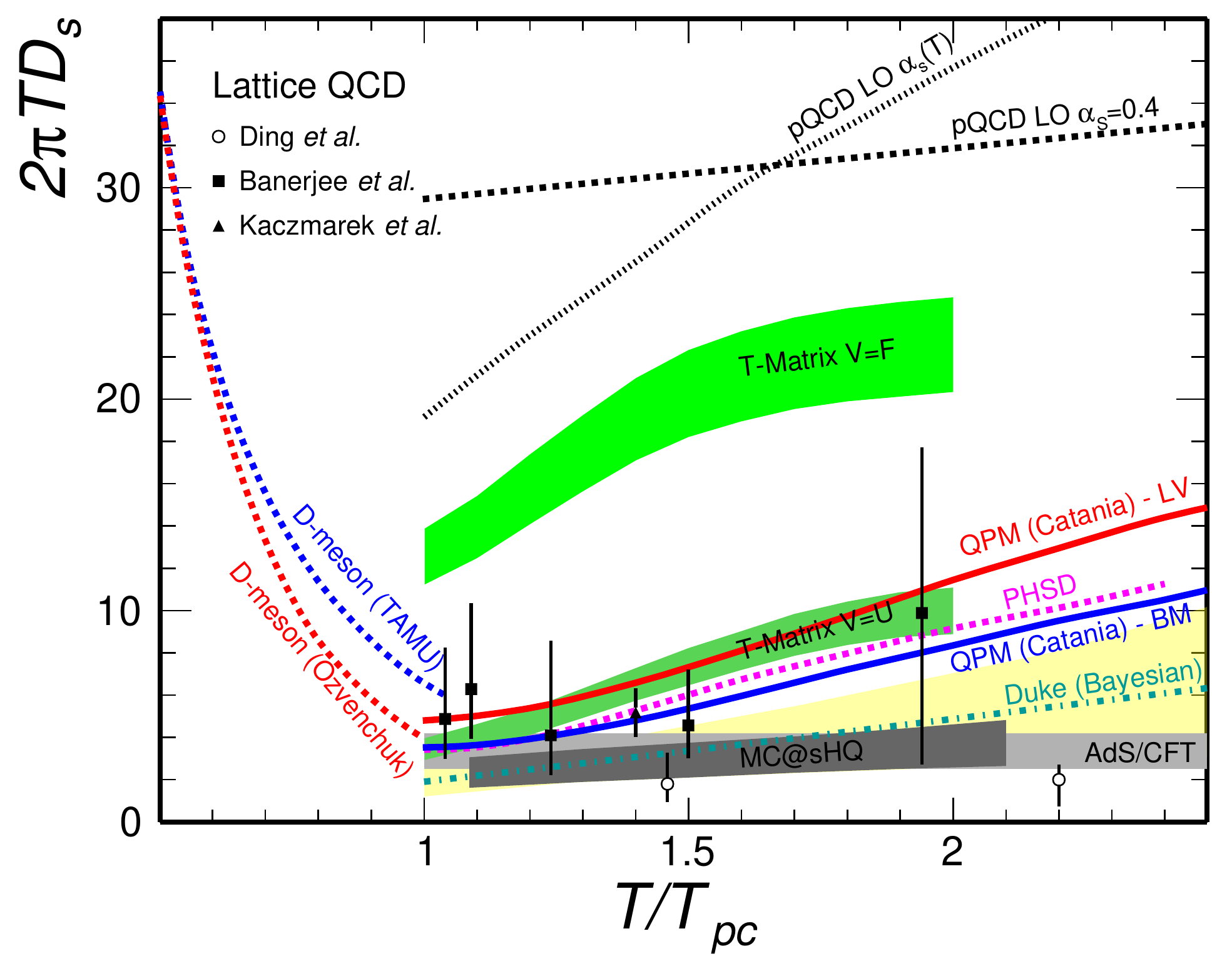}
\caption{Spatial diffusion coefficient ($2\pi TD_s$) of charm quark in the quark--gluon plasma calculated by different approaches versus the reduced temperature ($T/T_{pc}$). The lattice QCD calculations in the quenched approximation~\cite{Ding:2015ona,Banerjee:2011ra,Kaczmarek:2014jga} are compared with the estimations based on different models~\cite{Moore:2004tg,vanHees:2004gq,Das:2015ana,He:2014cla,Song:2015sfa,Andronic:2015wma,
Horowitz:2015dta,Xu:2017obm,He:2011yi,Tolos:2013kva}. The figure is from~\cite{Dong:2019byy}.}
\label{fig:Ds}
\end{figure}

The momentum diffusion coefficient $\kappa$ can be easily converted to the spatial one $D_s$ with the relation $\kappa=2T^2/D_s$. In recent years, the temperature dependence of the dimensionless quantity $2\pi TD_s$ has been estimated by a lot of theoretical frameworks, such as the lattice QCD (lQCD)~\cite{Ding:2015ona,Banerjee:2011ra,Kaczmarek:2014jga}, LO pQCD~\cite{Moore:2004tg,vanHees:2004gq}, QPM calculations~\cite{Das:2015ana}, \textit{T}-matrix~\cite{He:2014cla}, PHSD~\cite{Song:2015sfa}, MC$@_s$HQ~\cite{Andronic:2015wma}, Ads/CFT~\cite{Horowitz:2015dta}, duke (Bayesian analysis)~\cite{Xu:2017obm}, and hadronic matter~\cite{He:2011yi,Tolos:2013kva}, as shown in Figure \ref{fig:Ds}.

The estimates by the lQCD from the first principles provide a valuable reference for the model extractions of $2\pi TD_s$. As one can see, with relatively large uncertainties, the lQCD calculations in the quenched approximation give $D_s\sim$ 3.7--7.0~\cite{Kaczmarek:2014jga} over the temperature range from $T_{pc}$ to 2$T_{pc}$. However, it is difficult to extract meaningful information about the temperature dependence of $2\pi TD_s$ from the current lQCD results. Furthermore, except for the pQCD calculations at the leading-order which show obvious larger values than others, these extractions of $2\pi TD_s$ based on the recently developed models are consistent with the lQCD data, as well as previous studies presented in~\cite{Rapp:2018qla} which give $2\pi TD_s\sim$ 2--4 near the critical temperature. Although these calculations give different values of $2\pi TD_s$ versus $T/T_{pc}$, most estimations show that $D_s$ slightly increases with $T$. It implies that the interactions between a charm quark and the QCD medium have the strongest strength near the critical temperature. However, no direct evidence has been found in the experiment to verify this upward trend of $D_s$ so far because it is hard to find an observation only sensitive to the in-medium interactions at the late stage of the QGP evolution. Fortunately, the data-driven analysis utilizing Bayesian inferences seems to shed new light on this issue. The temperature and momentum dependence of $D_s$ has been extracted from the available experimental data ($R_{AA}$ and $v_2$ of a D meson both at the RHIC and LHC)~\cite{Xu:2017obm} based on the Duke--Langevin transport model, which indeed shows an upward trend of $2\pi TD_s$. More recently, this approach of Bayesian inference has been improved with the help of information field theory~\cite{Bialek:1996kd,Lemm:1999kd} in~\cite{Xie:2022fak}. Therefore, one can now extract model parameters without relying on an explicit form of parametrization, leading to a robust determination by such a model--data fit. 

\section{Production of Heavy-Flavour Jets in Heavy-Ion Collisions}
\label{sec:theory}

\subsection{Nuclear Modification Factors of Production Yields}
\label{sec:RAA}

To address the nuclear effect in relativistic heavy-ion collisions, the nuclear modification factor $R_{AA}$ is conventionally utilized to quantify the yield suppression of hadron/jet in A+A collisions per binary nucleon--nucleon collision relative to p+p~\cite{STAR:2003pjh},
\begin{eqnarray}
R_{AA}=\frac{1}{\left\langle N_{\rm bin}^{\rm AA} \right\rangle}\frac{d\sigma^{\rm AA}/\rm dydp_T}{d\sigma^{\rm pp}/\rm dydp_T}
\label{eq:raa}
\end{eqnarray}
where the scaling factor $\left\langle N_{\rm bin}^{\rm AA} \right\rangle$ denotes the number of binary nucleon--nucleon collisions in A+A~\cite{Miller:2007ri}. It has been observed that the values of $R_{AA}$ of hadrons and jets are smaller than one in nucleus--nucleus collisions both at the RHIC~\cite{STAR:2003wqp,PHENIX:2005nhb,PHENIX:2010bqp} and LHC~\cite{ATLAS:2018gwx,ATLAS:2014ipv}, and these measurements could be explained by the mechanism of partonic energy loss, which in turns serve as convincing evidence for the formation of QGP in such extremely hot and dense conditions. Meanwhile, the jet transport parameter $\hat{q}\equiv d\left\langle p^2_{\perp}\right\rangle/dL$~\cite{Baier:2002tc} representing the strength of in-medium partonic interactions could be extracted from the available $R_{AA}$ data by various theoretical models~\cite{JET:2013cls,Xie:2019oxg,Ru:2019qvz,JETSCAPE:2021ehl,Xie:2022fak}.

\begin{figure}[H]
\includegraphics[width=3.0in,height=2.7in,angle=0]{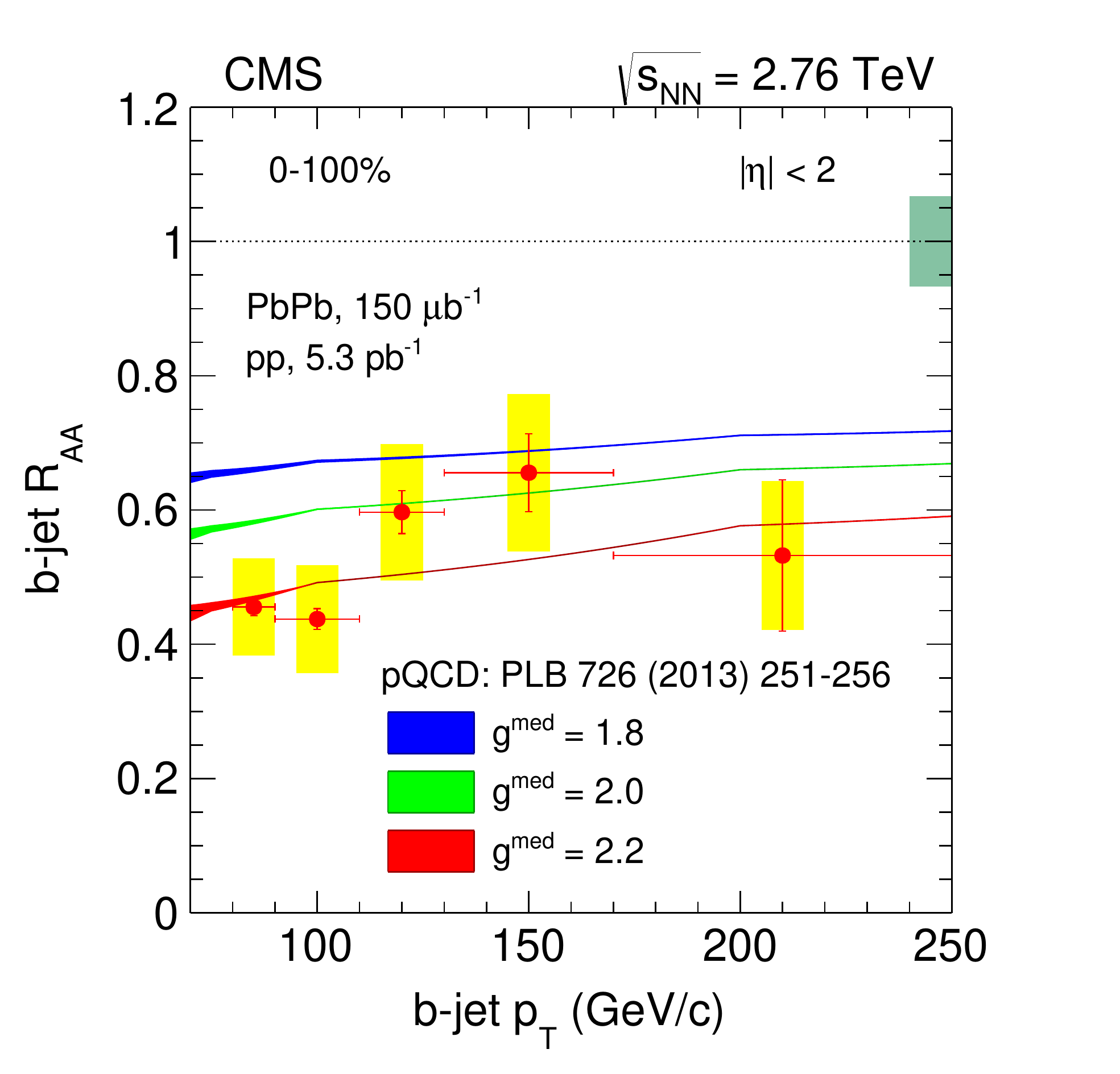}
\includegraphics[width=2.4in,height=2.5in,angle=0]{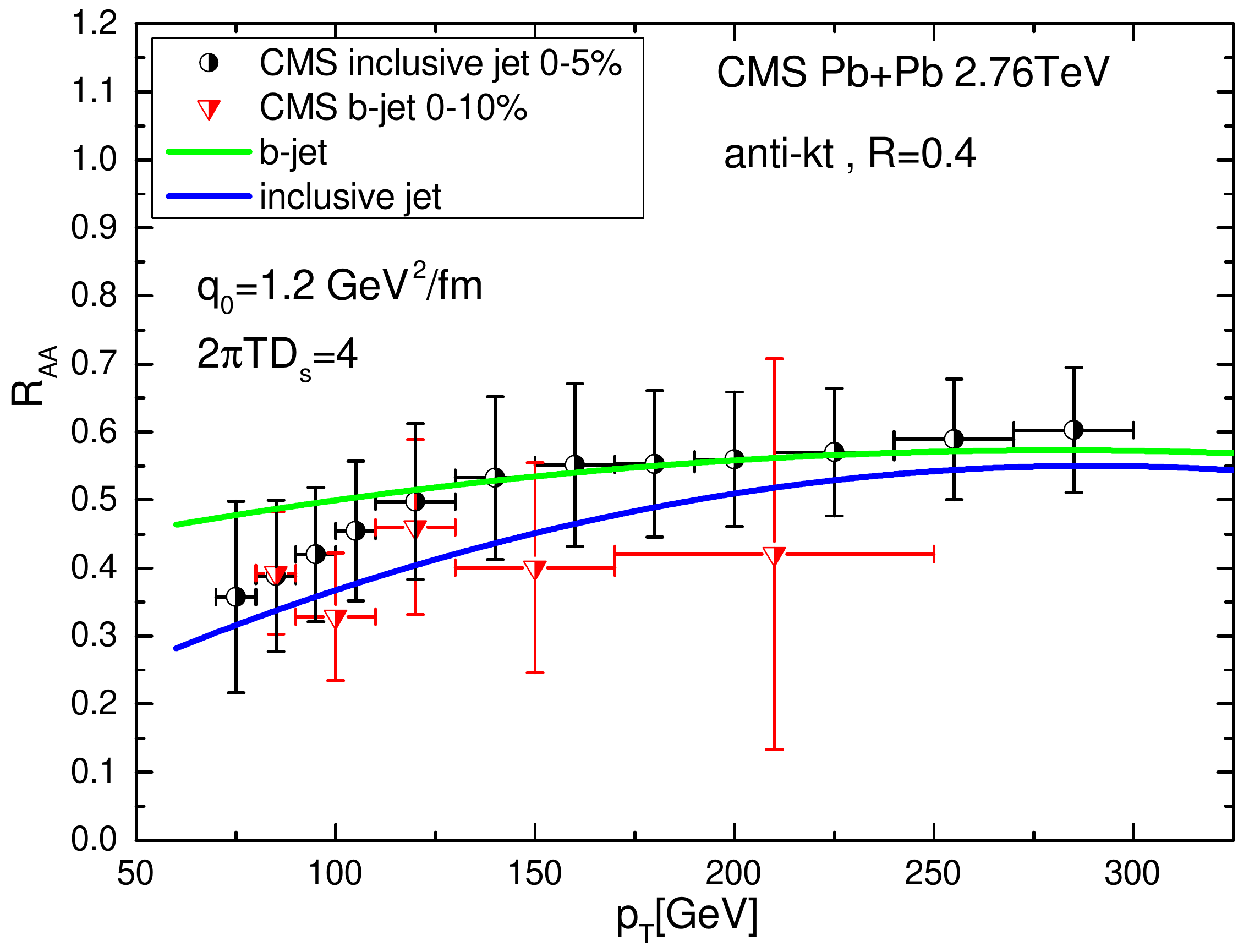}
\caption{\textbf{Left}: the measured nuclear modification factor $R_{AA}$ of the inclusive b-jet versus b-jet $p_T$ by the CMS collaboration in Pb+Pb at $\sqrt{s_{NN}}$~=~2.76 TeV for minimum bias collisions~\cite{Chatrchyan:2013exa}. \textbf{Right}: the comparison of $R_{AA}$ between the inclusive jets and b-jets versus jet $p_T$ in central Pb+Pb collisions at \mbox{$\sqrt{s_{NN}}$ = 2.76 TeV~\cite{Dai:2018mhw}}. The figures are from~\cite{Chatrchyan:2013exa,Dai:2018mhw}.}
\label{fig:raa}
\end{figure}

Additionally, to test the mass dependence of jet quenching, the $R_{AA}$ has also been used in a comparison of the yield suppression between heavy-flavour jets and inclusive jets. Benefiting from the fact that heavy-flavour jets are produced abundantly as the centre-mass energy increases in hadronic collisions at the LHC, the exploration of a heavy quark-tagged jet produced in heavy-ion collisions has gradually attracted much attention. The first experimental effort focused on the production of a b-jet was implemented by the CMS collaboration~\cite{Chatrchyan:2013exa} in 2013, as shown in the left plot of Figure~\ref{fig:raa}, where a b-jet is defined as jets containing at least one B hadron inside the jet-cone. The red points are the CMS data and the coloured bands are the theoretical calculations. This measurement accounts for the b-jet samples in minimum bias collisions (0--100\%). We note that even with large experimental uncertainties, the b-jet $R_{AA}$ slightly increases with jet $p_T$ and varies from 0.4 to 0.8. Significant suppression of the b-jet yield in Pb+Pb collisions at \mbox{$\sqrt{s_{NN}}$ = 2.76 TeV} relative to the p+p baseline was observed for the first time, which indicates that bottom quarks strongly interact with the hot/dense nuclear matter. Furthermore, within experiment uncertainties, the results were found to be consistent with the pQCD-based calculations conducted in~\cite{Huang:2013vaa} when the coupling factor $g^{med}$ varied from 1.8 to 2.2.
\vspace{-12pt}

To address the difference of the yield suppression between the b-jet and inclusive jet (mainly initiated by a massless light quark or gluon), a direct comparison of their $R_{AA}$ in the right plot of Figure~\ref{fig:raa} was presented by the SHELL approach, which applies a Langevin transport model to describe heavy quark propagation in the QGP~\cite{Dai:2018mhw} in central Pb+Pb collisions at $\sqrt{s_{NN}}$ = 2.76 TeV, as well as the next-to-leading order pQCD calculations matched with the parton shower effect for the p+p baseline~\cite{Gleisberg:2008ta,Frixione:2002ik}. In the model, the jet transport parameter $\hat{q}$ was extracted by the production of an identified hadron in A+A collisions~\cite{Ma:2018swx}, and then the spatial diffusion coefficient $D_s$ of heavy quarks can be determined by the D meson $R_{AA}$ data~\cite{Sirunyan:2017xss,ALICE:2018lyv}. The measured $R_{AA}$ of the inclusive jet with the centrality of \mbox{0--5\%~\cite{CMS:2016uxf}} and b-jet with 0--10\%~\cite{Chatrchyan:2013exa} are also illustrated in the plot of Figure~\ref{fig:raa}. Although the $R_{AA}$ of the b-jet seems to be slightly smaller than that of the inclusive jet, the CMS collaboration claims that no clear difference of $R_{AA}$ between the inclusive jet and b-jet was found, because the current uncertainties of the b-jet data are too large. However, the theoretical calculations in~\cite{Dai:2018mhw} suggest that b-jet $R_{AA}$ may be larger than inclusive jet $R_{AA}$, due to the ``dead-cone'' effect of the bottom quarks, which suppresses the medium-induced gluon radiation of massive heavy quarks within a cone $\theta\sim M/E$~\cite{Dokshitzer:2001zm}. A more precise measurement is necessary to resolve the tension between the experimental data and theoretical calculations. It is very exciting the fact that recently, the ATLAS collaboration reported preliminary results by simultaneously measuring the $R_{AA}$ of the inclusive jet and b-jet in 0--20\% Pb+Pb collisions at \mbox{$\sqrt{s_{NN}}$ = 5.02 TeV~\cite{ATLAS:2022fgb}}, which shows a clear weaker suppression of the b-jet, and the features can be described by the theoretical calculations~\mbox{\cite{Dai:2018mhw,Ke:2018tsh}}. Although the mass hierarchy of jet quenching at the particle level has been confirmed by a lot of experimental data~\cite{PHENIX:2022wim,STAR:2021uzu}, it is indisputable that the ATLAS measurement makes a crucial step towards finding the mass effect at the jet level. The comparison of the c- and b-jet $R_{AA}$ has been presented in~\cite{Li:2018xuv} with the SCET model~\mbox{\cite{Ovanesyan:2011kn,Sievert:2018imd}}, which shows no significant difference at $p_T>50$ GeV. More recently, some exploratory estimates indicate that the $R_{AA}$ of the c-jet may be stronger than that of the inclusive jet at higher jet $p_T$ due to their different constituents~\cite{Ke:2020nsm,wang:2022aaa}, an interesting finding to be investigated further in detail.

Beyond $R_{AA}$, another observation $I_{AA}$~\cite{Neufeld:2010fj} has also been utilized to study the yield suppression of b-jets tagged by $Z^0$ bosons in high-energy nuclear collisions~\cite{Wang:2020qwe}. Similar to $R_{AA}$, $I_{AA}$ is defined as follows,
\begin{eqnarray}
I_{AA}=\frac{1}{\left\langle N^{\rm AA}_{\rm bin} \right\rangle}\frac{\frac{d\sigma^{\rm AA}}{dp^{\rm jet}_{\rm T}}|_{p_{\rm T}^{\rm min}<p_{\rm T}^Z<p_{\rm T}^{\rm max}}}{\frac{d\sigma^{\rm pp}}{dp^{\rm jet}_{\rm T}}|_{p_{\rm T}^{\rm min}<p_{\rm T}^Z<p_{\rm T}^{\rm max}}}
\label{eq:IAA}
\end{eqnarray}

\begin{figure}[H]
\includegraphics[width=5.4in,height=2.2in,angle=0]{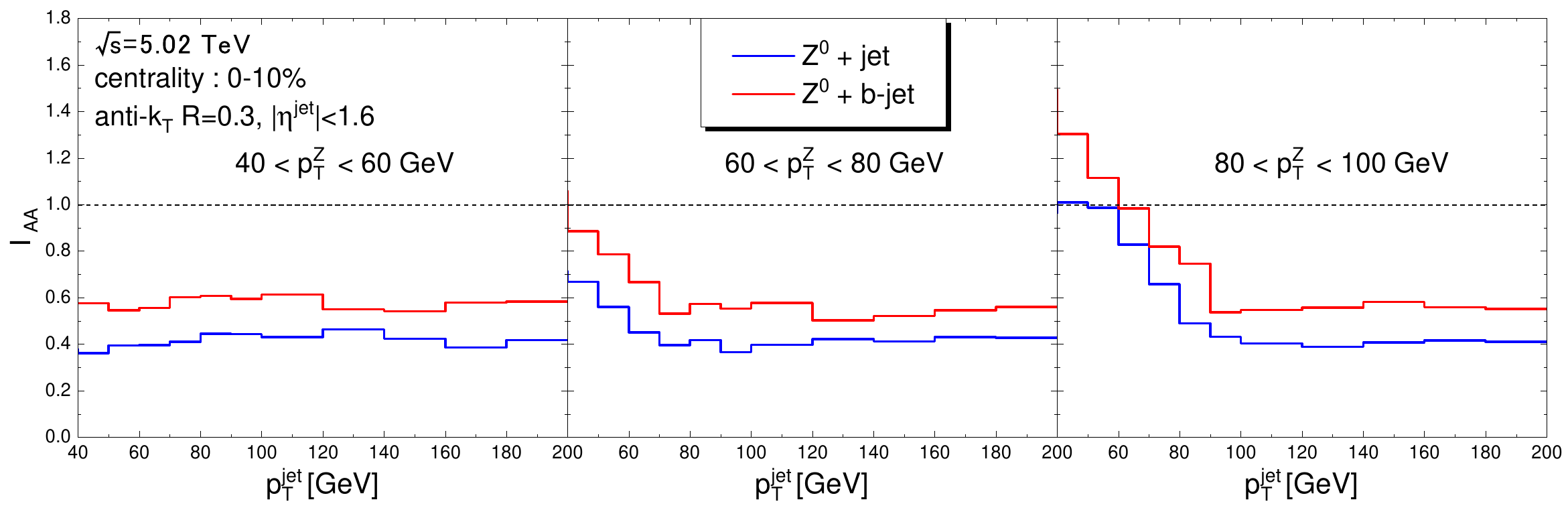}
\caption{{Nuclear modification factor} $I_{AA}$ as a function of the transverse momentum of the tagged jet within three $p_T^Z$ windows: 40--60, 60--80, 80--120~GeV in central 0--10\% Pb+Pb collisions at $\sqrt{s_{NN}}~=~5.02$~TeV~\cite{Wang:2020qwe}.}
\label{fig:IAA}
\end{figure}

One can see that $I_{AA}$ quantifies the yield variation of the jet in A+A collisions per binary nucleon--nucleon collision relative to the p+p baseline, after integrating the $Z^0$ boson $p_T$. It has been proposed that the associated production $Z^0\,+\,$b-jet may be helpful in addressing the mass dependence of the jet-quenching effect, since the $Z^0\,+\,$jet processes significantly exclude the contamination of gluon-initiated jets~\cite{Kartvelishvili:1995fr}. Thereby the comparison of $I_{AA}$ between the $Z^0\,+\,$jet and $Z^0\,+\,$b-jet can provide direct features of the mass effect of heavy quark jets compared to light-quark jets. Figure~\ref{fig:IAA} shows the comparisons of the calculated $I_{AA}$ of the $Z^0\,+\,$b-jet and $Z^0\,+\,$jet in central 0--10\% Pb+Pb collisions at $\sqrt{s_{NN}}$~=~5.02 TeV, where the jets are reconstructed by the anti-$k_T$ algorithm with a cone size R = 0.3 and pseudorapidity \mbox{$|\eta^{\rm jet}|<$ 1.6}. The calculations are presented within different $p_T$ windows of $Z^0$ bosons in the three panels, namely 40--60, 60--80, and 80--100 GeV. We note that the shapes of $I_{AA}$ are flat at the panel of $40<p_T^Z<60$ GeV but have downward trends at that of $80<p_T^Z<100$, and the $I_{AA}$ in the right panel even shows enhancement at\linebreak  $p_T^{\rm jet}$. This is because if one constrains the $Z^0$ meson momentum in the event selection, such as $80<p_T^Z<120$~GeV, then the selected jet distribution with $p_T$ would fall steeper at\linebreak $p_T^{\rm jet}<$ 80 GeV, which naturally leads to a relatively large value of the nuclear modification factors at \mbox{$p_T^{\rm jet}<$ 80 GeV}, even larger than one. Additionally, one observes that at each panel the model calculations show that $I_{AA}$ of the $Z^0\,+\,$b-jet is visibly higher than that of the $Z^0\,+\,$jet in nucleus--nucleus collisions, which indicates that the $Z^0$-tagged light-quark jets lose more energy than the $Z^0$-tagged b-jets traversing the QGP. These comparisons would be helpful to directly test the mass effect of jet energy loss in heavy-ion collisions at the LHC from a new perspective.

\newpage
\subsection{Transverse Momentum Imbalance}
\label{sec:XJ}

\begin{figure}[H]
\includegraphics[width=2.5in,height=2.4in,angle=0]{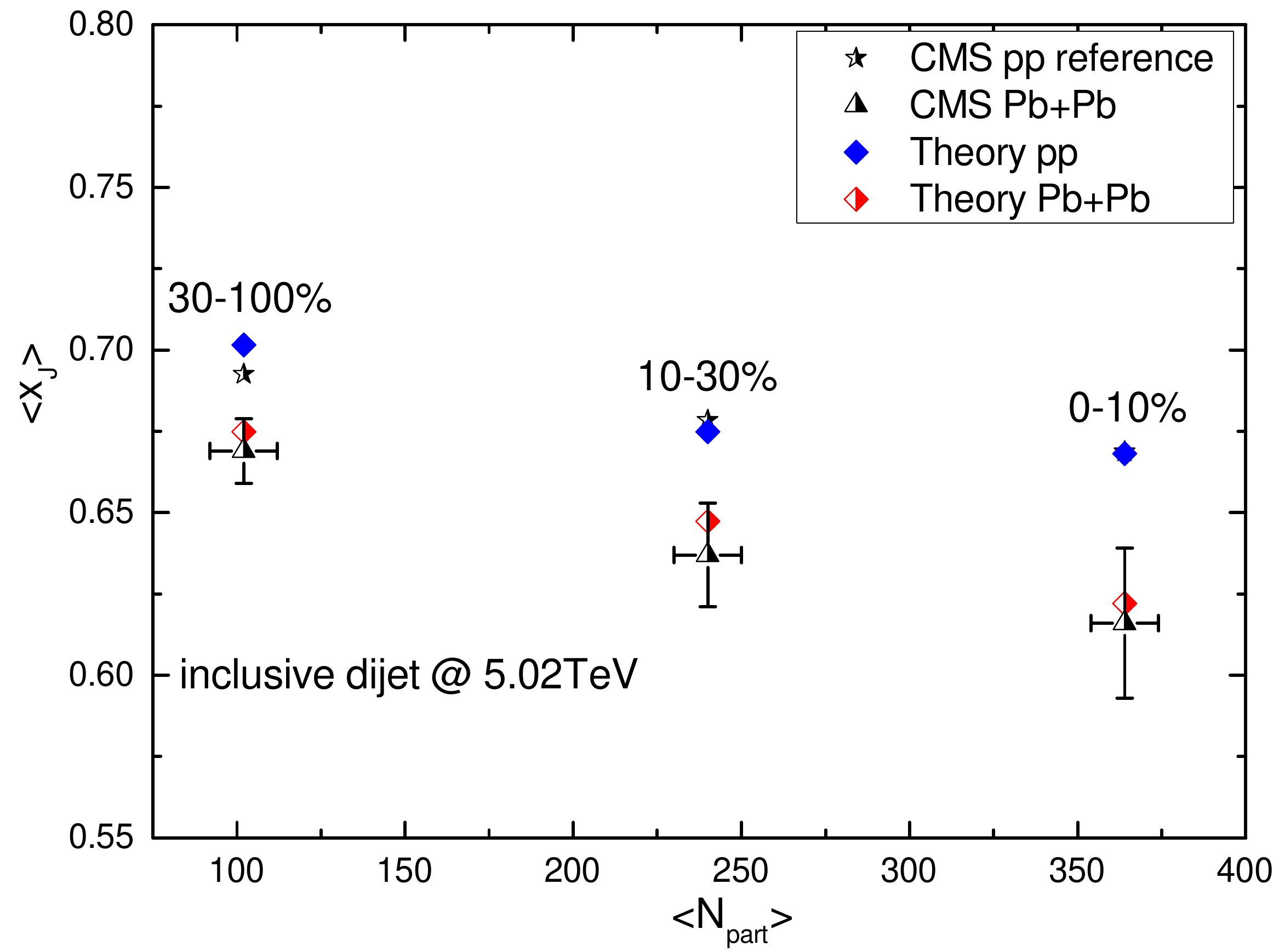}
\includegraphics[width=2.5in,height=2.4in,angle=0]{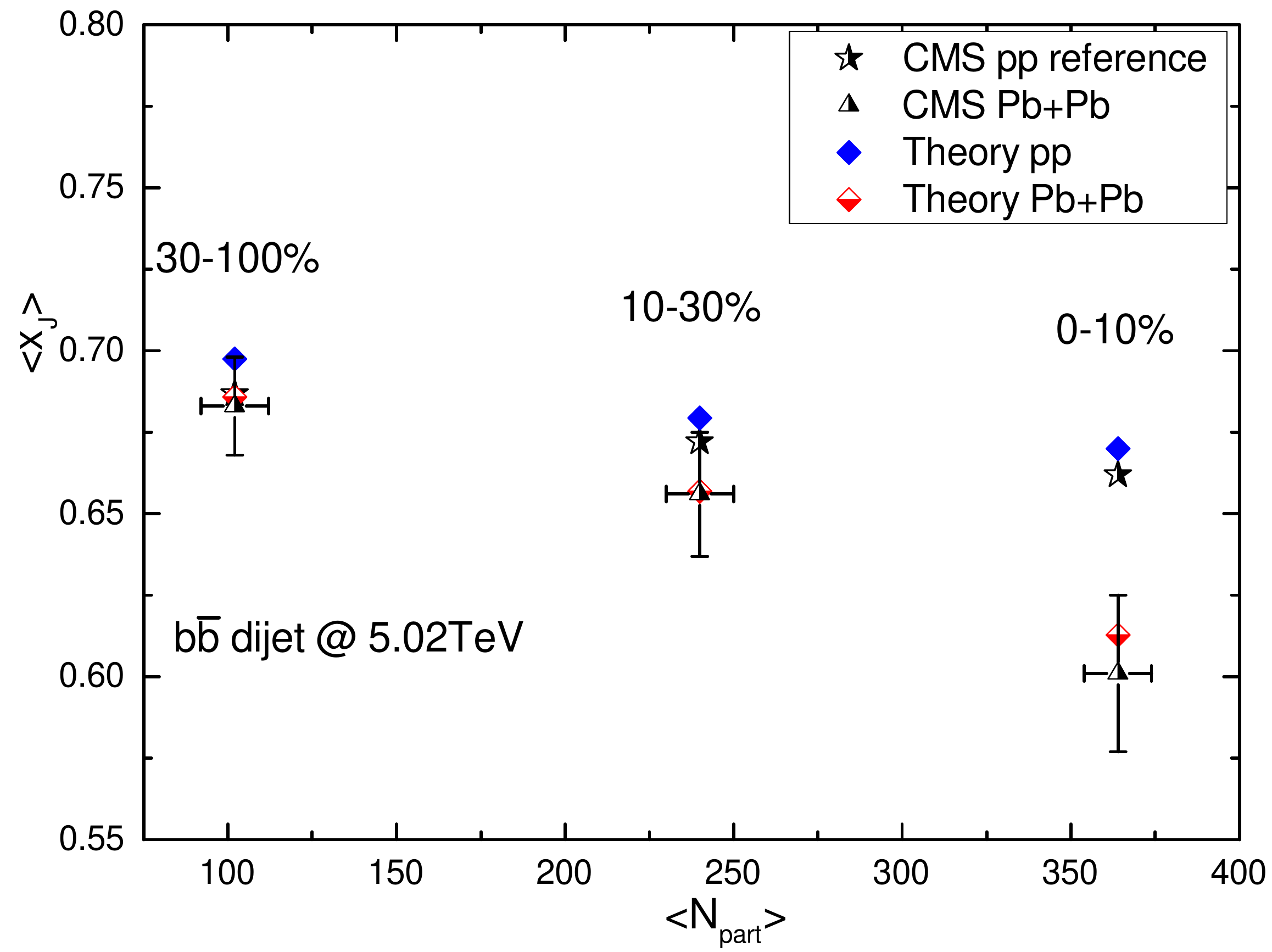}
\caption{Averaged $x_J$ value as a function of the number of participants calculated in p+p and Pb+Pb collisions $\sqrt{s_{NN}}$~=~5.02 TeV within different centrality bins compared with the experimental data, both for inclusive (\textbf{left}) and $b\bar{b}$ (\textbf{right}) dijets. Figures are from~\cite{Dai:2018mhw}.}
\label{fig:avexj}
\vspace{-0.2in}
\end{figure}

The transverse momentum imbalance ($x_J=p_{\rm T,2}/p_{\rm T,1}$) is another useful observation, describing the momentum asymmetry of the dijet system in the transverse plane, where $p_{\rm T,1}$ and $p_{\rm T,2}$ denote the leading and sub-leading jet $p_T$. It is noted that in the fixed-leading-order QCD calculations the two outgoing hard partons should be strictly back-to-back in the transverse ($x_J$~=~1), but the higher-order corrections and vacuum shower may break the symmetry which leads to $x_J<$ 1. In heavy-ion collisions, the smaller $x_J$ of the $\gamma$+jet~\cite{CMS:2017ehl} and $Z^0$+jet~\cite{CMS:2017eqd} systems have been observed in Pb+Pb collisions compared to p+p, which results from the energy loss of the tagged jet. The CMS collaboration reports the measurement on $x_J$ of the inclusive and $b\bar{b}$ dijets in Pb+Pb collisions at \mbox{$\sqrt{s_{NN}}$~=~5.02 TeV~\cite{Sirunyan:2018jju}}. In their measurements, the biggest challenge was how to select the $b\bar{b}$ dijet events initiated by the hard heavy-quark pairs, because it is crucial to address the mass effect by directly comparing such heavy-quark dijets with inclusive dijets. On the theoretical side, the production mechanisms of heavy quarks can be categorized into three classes: flavour creation (FCR), flavour excitation (FEX), and gluon splitting (GSP)~\cite{Nason:1989zy,Beenakker:1990maa,Mangano:1991jk,Norrbin:2000zc}, only FCR represents the dijets initiated by heavy-quark pairs originating from the hard process. The CMS collaboration suggests a strategy to separate the FCR processes by selecting $b\bar{b}$ dijets that have a large opening angle ($\vert \Delta \phi \vert > 2\pi/3$) in azimuth, which could significantly suppress the contributions of the other two. This method has also been used in theoretical studies~\cite{Huang:2015mva,Dai:2018mhw,Kang:2018wrs}. Figure~\ref{fig:avexj} shows a comparison of the averaged $x_J$ of the inclusive and $b\bar{b}$ dijets in both p+p and Pb+Pb collisions with different centrality bins at $\sqrt{s_{NN}}$~=~5.02 TeV, as well as the experimental data~\cite{Sirunyan:2018jju}, where $\left\langle x_J \right\rangle$ was estimated as follows.
\vspace{-6pt}
\begin{eqnarray}
\left\langle x_J \right\rangle = \frac{1}{\sigma}\int_{0}^{1}\frac{d\sigma}{dx_J}dx_J
\end{eqnarray}

{The} black triangle points are the CMS data in Pb+Pb collisions, and the black star points are the p+p reference used in their measurements. The blue and red rhombus points are the theoretical calculations, while the p+p reference is provided by the Monte Carlo event generator SHERPA~\cite{Gleisberg:2008ta} which matches the next-to-leading order QCD matrix elements and the parton shower effect in a vacuum~\cite{Gleisberg:2008fv,Schumann:2007mg}. The $\left\langle x_J \right\rangle$ points of the inclusive (left panel) and $b\bar{b}$ (right panel) dijets are listed within three centrality bins which correspond to the different numbers of the participant in Pb+Pb collisions. In Figure~\ref{fig:avexj}, the theoretical calculations based on the SHELL model~\cite{Dai:2018mhw} show an overall decrease in $\left\langle x_J \right\rangle$ in Pb+Pb collisions relative to the p+p baseline both for the inclusive and $b\bar{b}$ dijets, consistent with the CMS data and indicates that the asymmetry between these two leading jets is amplified in A+A collisions. The reduction in $\left\langle x_J \right\rangle$ is centrality-dependent since the in-medium interaction is sensitive to the temperature and size of the QGP. It is even more important that the calculations show that the decrease in $\left\langle x_J \right\rangle$ of the $b\bar{b}$ dijets is slightly smaller than that of the inclusive dijets within the same centrality bins. These results suggest that dijets initiated by bottom quarks may suffer smaller energy loss compared to those initiated by light quarks or gluons. Furthermore, another study on the $b\bar{b}$ dijet in heavy-ion collisions~\cite{Kang:2018wrs} proposed that the invariant mass $m_{jj}$ of the dijet system could be a novel observation sensitive to mass effects of jet quenching.

\begin{figure}[H]
\includegraphics[width=2.6in,height=3.2in,angle=0]{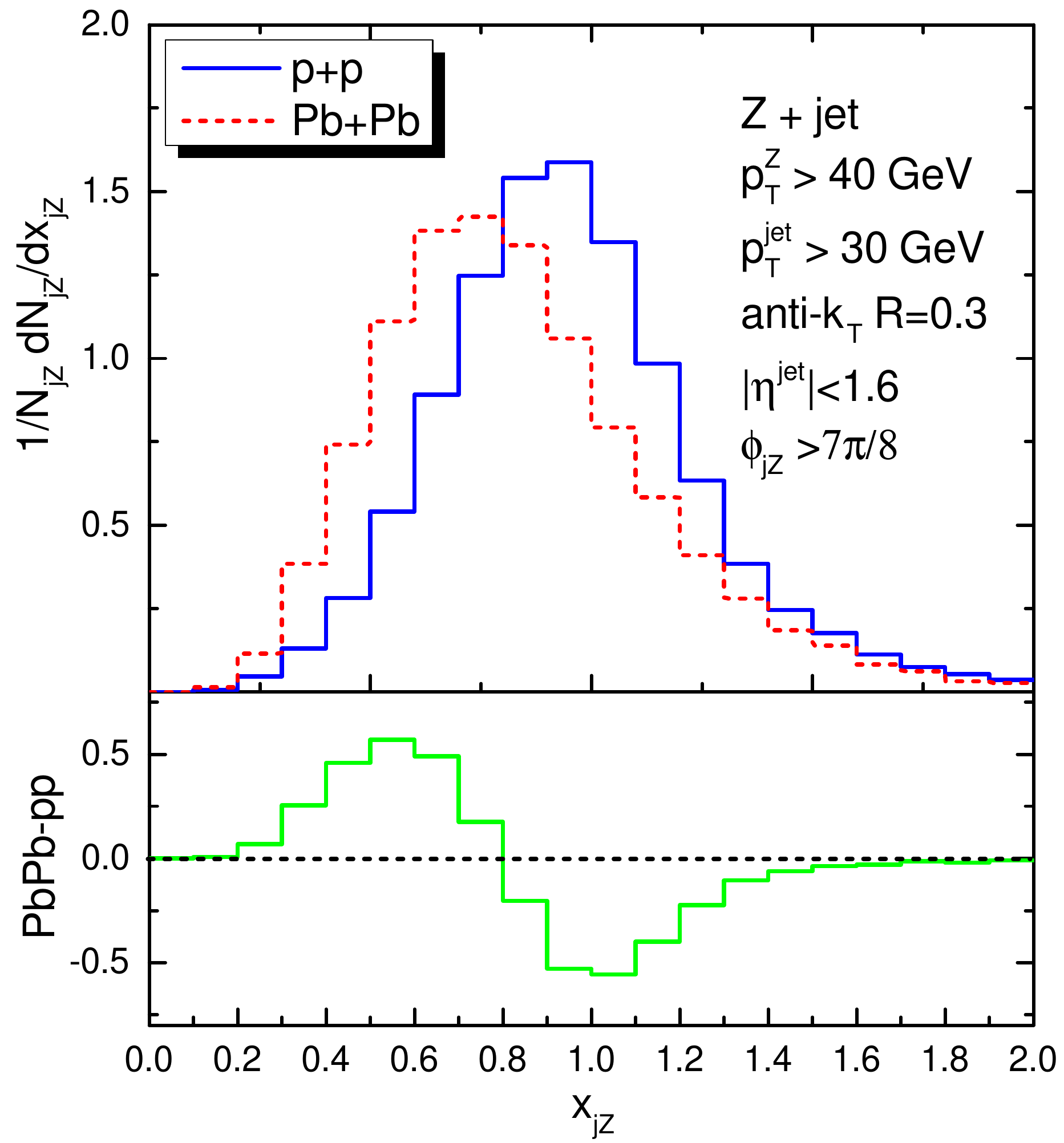}
\includegraphics[width=2.6in,height=3.2in,angle=0]{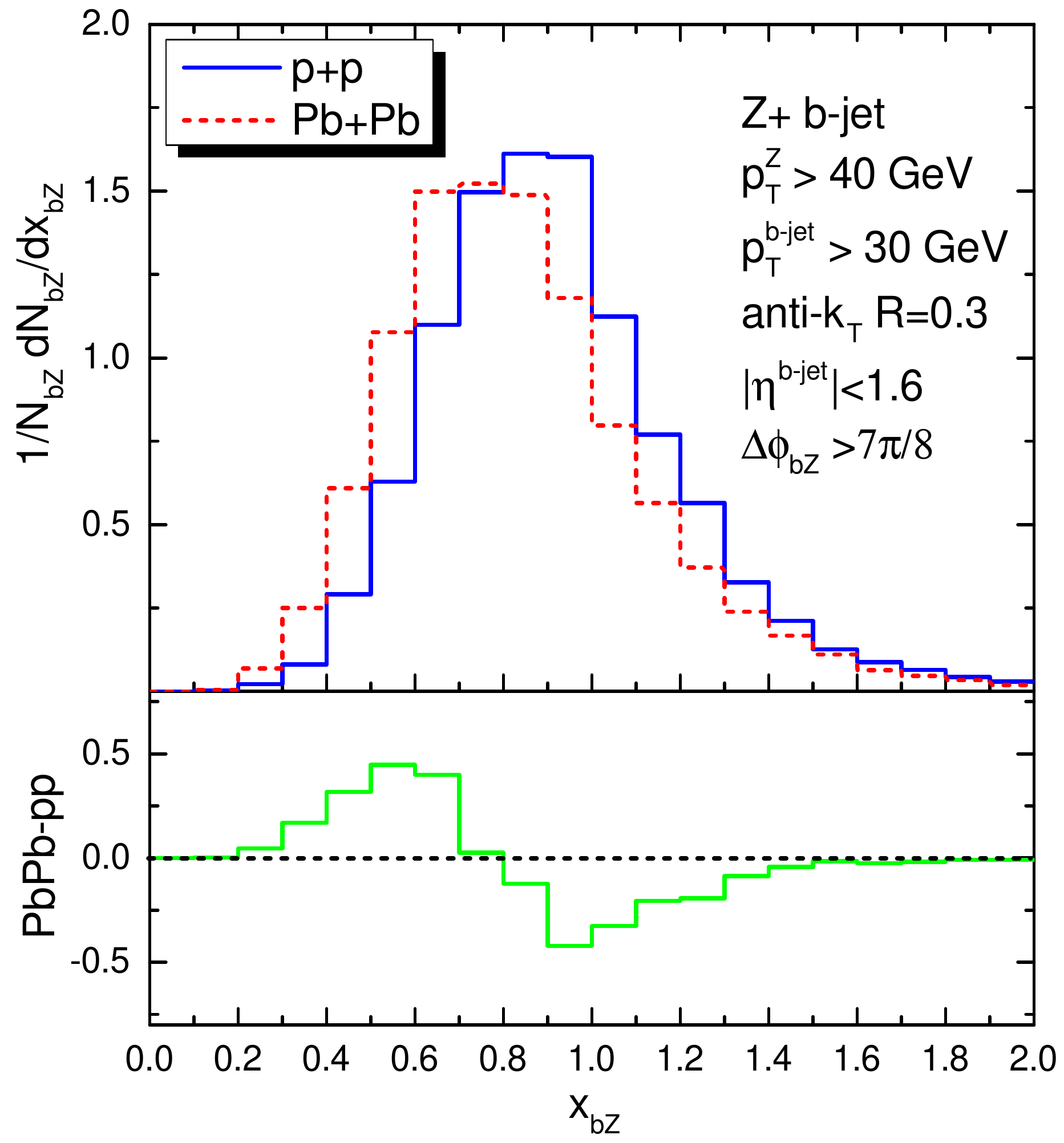}
\caption{{Distributions} of $x_{jZ}$~($x_{bZ}$) of $Z^0\,+\,$jet (\textbf{left}) and $Z^0\,+\,$b-jet (\textbf{right}) both in p+p and 0--10\% Pb+Pb  collisions at $\sqrt{s_{NN}}$~=~5.02~TeV. The differences of $x_{jZ}$~($x_{bZ}$) distributions in p+p and Pb+Pb collisions (green line) are also shown in the lower panels. Figures are from Ref.~\cite{Wang:2020qwe}.}
  \label{fig:xjib}
\end{figure}

In addition to the dijet system, the transverse momentum imbalances of the $Z^0\,+\,$jet ($x_{jZ}=p_T^{\textit{\rm jet}}/p_T^Z$) and $Z^0\,+\,$b-jet ($x_{bZ}=p_T^{\textit{\rm b-jet}}/p_T^Z$) have also been investigated~\cite{Wang:2020qwe}. It was found that the $Z^0$-tagging requirement considerably decreased the contribution of gluon-jets by 40\% in $Z^0\,+\,$jets compared to the dijet sample, especially at a lower jet $p_T$. The comparison of the medium modification on the $x_J$ of $Z^0\,+\,$jet and $Z^0\,+\,$b-jet may be suitable to address the mass effect of jet quenching. Figure~\ref{fig:xjib} shows the distributions of the $x_{jZ}$ (left) and $x_{bZ}$ (right) both in p+p and 0--10\% Pb+Pb collisions at $\sqrt{s_{NN}}$~=~5.02~TeV. In the calculations, the selected $Z^0$ bosons are required to have $p_T^Z>$ 60 GeV. The tagged jets (b-jets) are reconstructed with the anti-$k_T$ algorithm with a cone-size R = 0.3 and pseudorapidity $|\eta^{\rm jet}|<$ 1.6, required to have \mbox{$p_T^{\rm jet}>$ 30 GeV}. In particular, to guarantee that the $Z^0$ bosons and the tagged jets are back-to-back in the transverse plane, the $Z^0\,+\,$jet or $Z^0\,+\,$b-jet pairs are required to have a large opening angle in azimuth, $\Delta\phi_{jZ}(\Delta\phi_{bZ})>7\pi/8$. The differences of $x_{jZ}$~($x_{bZ}$) distributions in p+p and Pb+Pb collisions are also shown in the lower panels. Due to the jet energy loss, the $x_{jZ}$ and $x_{bZ}$ distributions shift towards smaller $x_J$ values in Pb+Pb collisions relative to p+p. Furthermore, one can find in the lower panel that the variations of $x_{bZ}$ are slightly smaller than that of $x_{jZ}$. More intuitive comparisons between the averaged $x_{jZ}$ and $x_{bZ}$ are listed in Table \ref{tab:axjz}. Within the statistical errors, the results show that $\Delta \left\langle x_{jz} \right\rangle \sim$ 0.136 is considerably larger than $\Delta \left\langle x_{bz} \right\rangle \sim$ 0.092, consistent with the expectation that bottom jets lose less energy than light-quark jets.
\vspace{-6pt}
\begin{table}[H]
\caption{The averaged $x_J$ of $Z^0$ + jet and $Z^0$ + b-jet both in p+p  and Pb+Pb collisions at \mbox{$\sqrt{s_{NN}}=5.02$~TeV}, as well as their variations $\Delta x_{J}={\left\langle x_{J} \right\rangle }_{\rm pp}-{\left\langle x_{J} \right\rangle }_{\rm PbPb}$. The statistical errors of $x_J$ in the simulations are also presented. Table is from Ref.~\cite{Wang:2020qwe}.}\label{tab:axjz}
\begin{tabular}{p{4.0cm}<{\centering}p{4.3cm}<{\centering}p{4.3cm}<{\centering}}
  \toprule
  & \boldmath{$Z^0$} \textbf{+ jet} & \boldmath{$Z^0$} \textbf{+ b-jet} \\
  \midrule
  ${\left\langle x_{J} \right\rangle }_{pp}$ & 0.987 $\pm$ 0.0047 & 0.941 $\pm$ 0.0056 \\
  \midrule
  ${\left\langle x_{J} \right\rangle }_{PbPb}$ & 0.851 $\pm$ 0.0061 & 0.849 $\pm$ 0.0064 \\
  \midrule
  $\Delta \left\langle x_{J} \right\rangle $ & 0.136 $\pm$ 0.0108 & 0.092 $\pm$ 0.012 \\
  \bottomrule
\end{tabular}
\end{table}

\subsection{Angular Correlation}
\label{sec:corre}

\begin{figure}[H]
\includegraphics[width=2.2in,height=2.4in,angle=0]{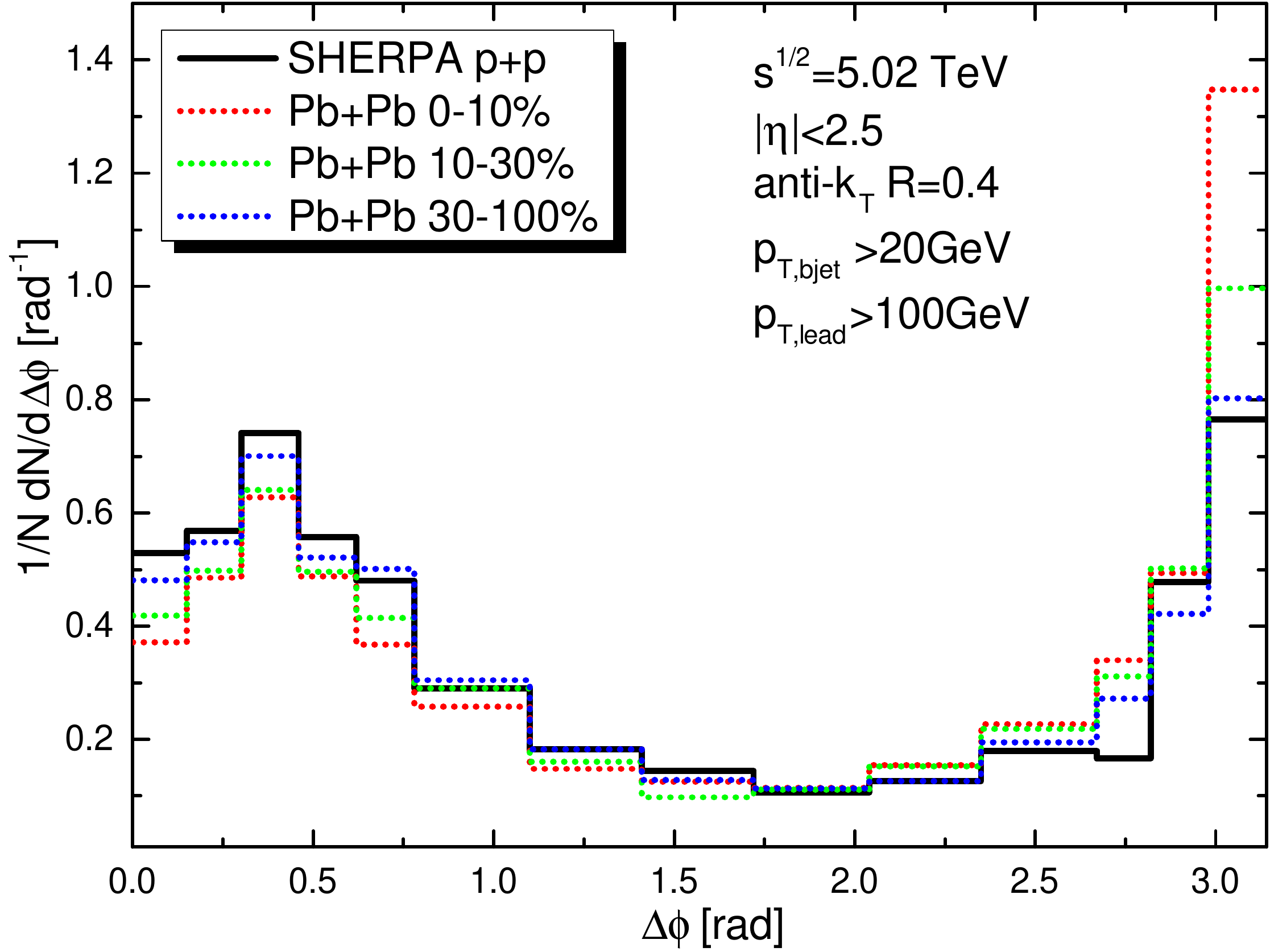}
\includegraphics[width=2.2in,height=2.45in,angle=0]{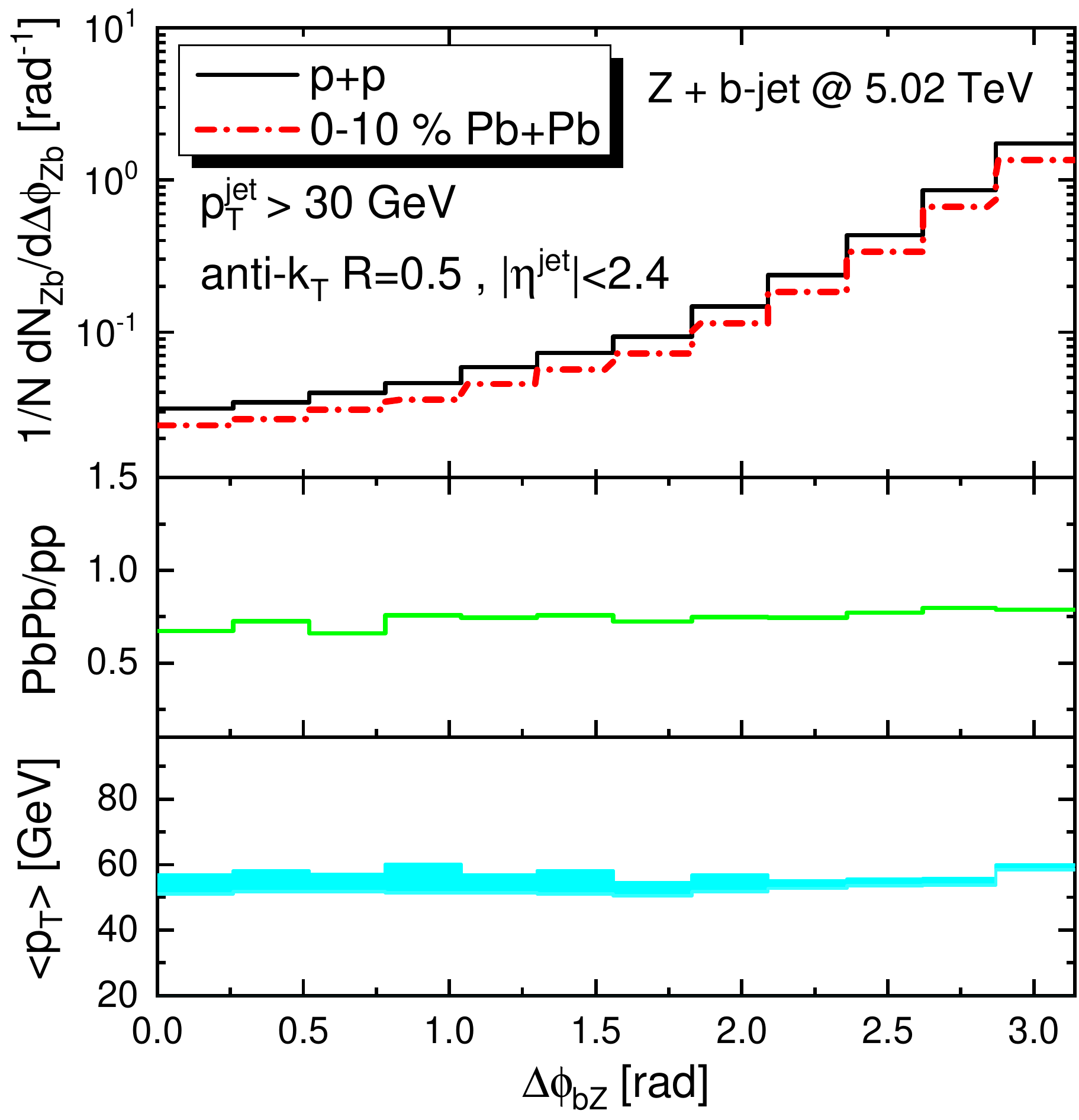}
\caption{\textbf{Left}: normalized azimuthal angular distributions of $b\bar{b}$ dijet system in p+p and Pb+Pb collisions at $\sqrt{s_{NN}}$~=~5.02 TeV. Results for different centrality bins, 0--10\%, 10--30\%, 30--100\%, are presented. \textbf{Right}: {the azimuthal} angular distribution of $Z^0\,+\,$b-jet in p+p and 0--10\% Pb+Pb collisions at $\sqrt{s_{NN}}$~=~5.02 TeV in the upper panel, while the ratio of PbPb/pp (green solid line) was shown in the middle panel and the averaged b-jet $p_T$ (blue band) in the lower panel. Figures are from Refs.~\cite{Wang:2018gxz,Wang:2020qwe}.}
\label{fig:angular}
\end{figure}

Jet angular correlations, such as $\Delta \phi$ distribution of dijets~\cite{Mueller:2016gko,Jia:2019qbl} and $\gamma/Z^0$ +\linebreak jet~\mbox{\cite{Luo:2018pto,Zhang:2018urd}}, are useful observable to address the medium-induced transverse momentum effect. In this context, estimating the medium modification on the angular distribution of heavy quark dijets in nucleus--nucleus collisions may also be of interest from the theoretical point of view. As shown in the left plot of Figure~\ref{fig:angular}, medium modification of the azimuthal angular correlations~($\Delta\phi=|\phi_{b1}-\phi_{b2}|$) of the $b\bar{b}$ dijet system in Pb+Pb collisions with different centralities at $\sqrt{s_{NN}}$~=~5.02 TeV are calculated~\cite{Wang:2018gxz}. One can observe suppression at $\Delta\phi\sim$0 and enhancement at $\Delta\phi\sim\pi$ in Pb+Pb collisions compared to the p+p, and the modifications are centrality dependent. Since the distributions are self-normalized, it implies that $b\bar{b}$ dijets with a larger opening angle (back-to-back) suffer relatively weaker yield suppression compared to that with a smaller one (collinear). It can be noted that the main contribution of $b\bar{b}$ dijet production at smaller $\Delta\phi$ is from the GSP process while larger $\Delta\phi_{bb}$ from the FCR process. The two b-jets from the former process share the energy of the gluon and then usually have lower $p_T$ than that from the latter process. As a result, the yield at the smaller $\Delta\phi$ region is more sensitive to the selection cut $p_T^{\rm jet}>$ 20 GeV. Actually, in another study on the angular correlations of $Z^0\,+\,$b-jet~\cite{Wang:2020qwe}, it's found that initial average b-jet $p_T$ distribution versus $\Delta\phi$ play a critical role, as shown in the right plot of Figure~\ref{fig:angular}. We see that the ratio of PbPb/pp in the middle panel is flat, and the average b-jet $p_T$ distribution is also flat. It's reasonable to guess that in Pb+Pb the azimuthal angle between b-jet and $Z^0$ has not been modified compared to p+p, and the overall suppression occurs at whole $\Delta\phi_{\rm bZ}$ region. Of course, we can imagine that it is more difficult for high-$p_T$ ($>$30 GeV) jets to be significantly deflected by the scattering with thermal parton.

To probe the angular deflection caused by the in-medium $p_T$-broadening,  observables accessible to lower $p_T$ region are needed. For this reason, it's proposed in Ref.~\cite{Wang:2021jgm} that the heavy-flavour meson tagged by direct photon ($\gamma+$HF) may provide a promising channel, with several advantages: (1) the transverse momentum resolution of $D^0$ meson can be low down to $\sim$1 GeV~\cite{ALICE:2018lyv} where the angular deflection is significant, (2) the photon gauges the initial momentum of heavy quarks, therefore, it's easy to quantify the direction change, (3) the selection bias effect can be suppressed by constraining the photon energy \ to cite{Cunqueiro:2021wls}. In this way, the considerable angular de-correlations between the heavy quarks and photons are predicted both in central Au+Au collisions at the RHIC and Pb+Pb collisions at the LHC. Furthermore, by constructing the 2-dimensional ($\Delta\phi, x_J$) correlation diagram of $\gamma+$HF, it's argued that the two aspects of jet quenching, energy loss, and $p_T$-broadening, can be well displayed simultaneously. Additionally, it's noted that another measurement on the angular correlations of $D^0$+hadron in Au+Au collisions at $\sqrt{s_{NN}}$~=~200 GeV may reflect the medium modification of the charm+jet correlation in the $\eta-\phi$ plane~\cite{STAR:2019qbf}, that awaits further detailed investigations.

\subsection{Radial Profile}
\label{sec:radial}

\begin{figure}[H]
\includegraphics[width=2.5in,height=2.7in,angle=0]{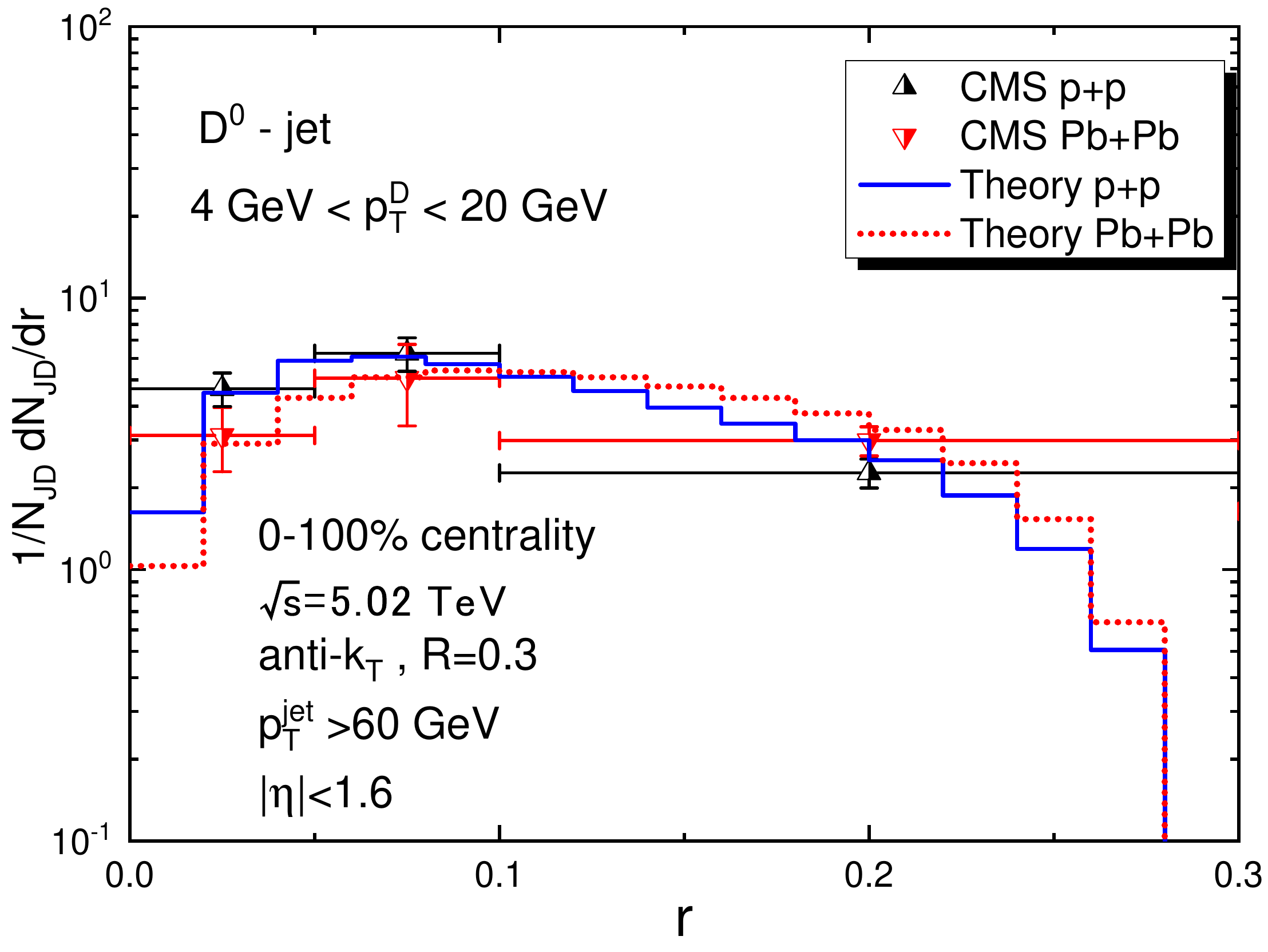}
\includegraphics[width=2.5in,height=2.7in,angle=0]{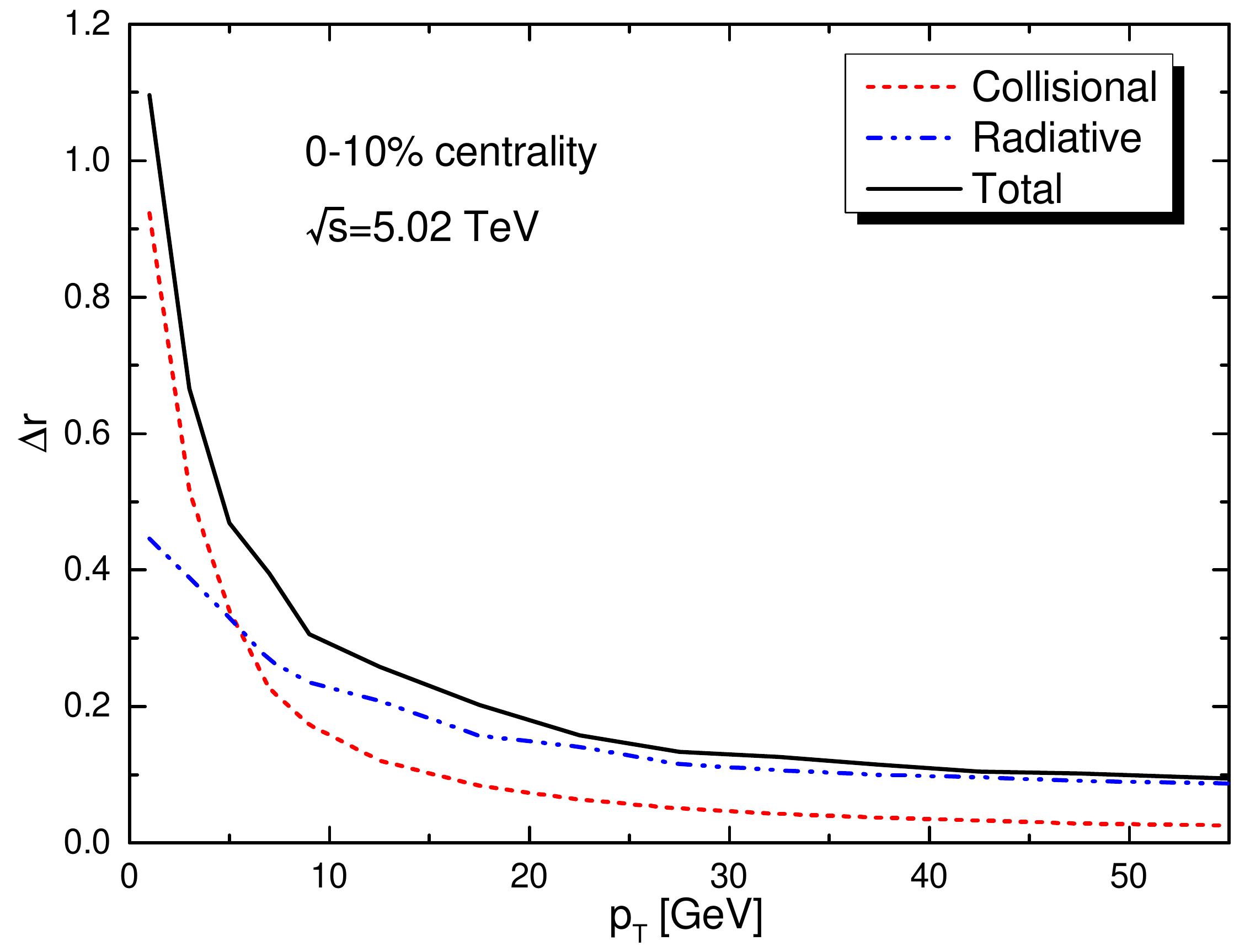}
\caption{\textbf{Left}: radial profile of D-jet in p+p and Pb+Pb collisions at $\sqrt{s_{NN}}$~=~5.02 TeV. \textbf{Right}: angular deviation of charm quark as a function of initial $p_T$. Figures are from Refs.~\cite{Wang:2019xey,Wang:2020bqz}.}
\label{fig:radial}
\end{figure}

The radial profile of the heavy-flavour jet represents the distribution of the angular distance $r=\sqrt{(\phi_{\rm Q}-\phi_{\rm jet})^2+(\eta_{\rm Q}-\eta_{\rm jet})^2}$ between the heavy-flavour meson and the jet-axis in the $\eta-\phi$ plane. Systematic studies with a focus on the radial profiles of D-jet and B-jet in heavy-ion collisions are performed in Refs.~\cite{Wang:2019xey,Wang:2020bqz}. As shown in the left panel of Figure~\ref{fig:radial}, the model calculated radial profiles of D-jets both in p+p and 0--100\% Pb+Pb collisions at $\sqrt{s_{NN}}=$ 5.02 TeV compared to the CMS measurements~\cite{Sirunyan:2019dow}. The black and red triangle points represent the measured data. The D-jets are reconstructed with anti-$k_T$ algorithm with R = 0.3 and $|\eta^{\rm jet}|<$ 1.6. All selected D-jets must satisfy $p_T^{\rm jet}>$ 60 GeV and contain at least one $D^0$ meson in jet-cone with $4<p_T^D<20$ GeV. The blue solid line is the p+p baseline provided by SHERPA~\cite{Gleisberg:2008ta}, and the red dashed line denotes the calculations based on the SHELL model. One can observe that the model calculations show the radial profile of D-jets in Pb+Pb collisions shifts towards larger radii relative to that of p+p, which is consistent with the diffusion trend observed by the CMS collaboration. These results show a clear physics picture, that charm quarks change their moving direction when scattering with the thermal partons in the hot and dense QCD matter. The studies argue that the diffusion behavior of D meson is closely related to the $p_T$-broadening when charm quarks scatter with the thermal partons in the medium. It should be noted that in such an estimate the jets are required to have $p_T>60$ GeV while D meson $p_T<20$ GeV, which makes that the higher $p_T$ jets can be viewed as a reference to probe the moving direction changes of charm quarks. It is found that the angular deviation $\Delta r=\sqrt{(\phi_c^f-\phi_c^i)^2+(\eta_c^f-\eta_c^i)^2}$ of charm quarks from their initial position in the $\eta-\phi$ plane is $p_T$ dependent, as shown in the right plot of Figure~\ref{fig:radial}. The charm quarks with lower $p_T$ are more likely to change their traveling direction via the in-medium scattering, and this feature also explains why no visible modification is observed in the CMS data for $p_T^D>20$ GeV~\cite{Sirunyan:2019dow}. The angular deviation at lower $p_T$ (\mbox{$<$5 GeV}) is dominated by elastic scattering, whereas at higher $p_T$ by inelastic reactions. These investigations may cast light on the in-medium energy loss mechanisms and constrain the transport coefficients of heavy quarks from a new perspective. We notice that a preliminary result of the D-jet radial profile in Au+Au collisions at \mbox{$\sqrt{s_{NN}}=$ 200 GeV} has been reported by the STAR collaboration in Ref.~\cite{Roy:2022yrw}. This result shows a similar diffusion effect of charm quark in jets in mid-central 10--40\% collisions.

To test the mass effect reflected in the radial profile, an additional comparison of the medium modification between D-jet and B-jet has been presented in Refs.~\cite{Wang:2020ffj,Wang:2020ukj}, where an inverse modification pattern on the radial profile of B-jets compared to D-jets is observed. The jet quenching effect seems to narrow the jet radial profiles of B-jets while broadening those of D-jets. It's demonstrated that the selection bias effect~\cite{Renk:2012ve} in A+A collisions may play a pivotal role. Heavy quark jets with higher $p_T$ have narrower initial radial distributions, and would naturally lead to narrower modifications when they fall into the lower $p_T$ domain due to jet energy loss. This reveals the fact that the final-state modification of the jet is not only influenced by the pure medium effect, but also by the other factors, such as the initial spectra and the selection bias~\cite{Cunqueiro:2021wls}.

\subsection{Fragmentation Function}

The jet fragmentation function $D(z)=(1/N_{\rm jet})dN_{\rm ch}(z)/dz$ is one of the most well-explored jet substructure observable~\cite{Sjostrand:1986hx,Webber:1983if,Procura:2009vm}, which usually refers to the longitudinal momentum distribution of charged hadrons inside the jet-cone~\cite{CMS:2014jjt,ATLAS:2014dtd,ATLAS:2017nre,ATLAS:2019dsv}. For heavy-flavour jets, the corresponding observable is the longitudinal momentum distribution of heavy-flavour mesons in jets, defined as in~\cite{ALICE:2019cbr}.
\begin{eqnarray}
D(z_{||})=\frac{1}{N_{\rm jet}}\frac{dN_{\rm HQ}(z_{||})}{dz_{||}}, \quad {\rm where} \
z_{||}=\frac{\vec{p}_{\rm HQ} \cdot \vec{p}_{\rm jet}}{\vec{p}_{\rm jet} \cdot \vec{p}_{\rm jet}} \,\,  .
\label{eq:dz}
\end{eqnarray}

\begin{figure}[H]
\includegraphics[width=5.2in,height=4.8in,angle=0]{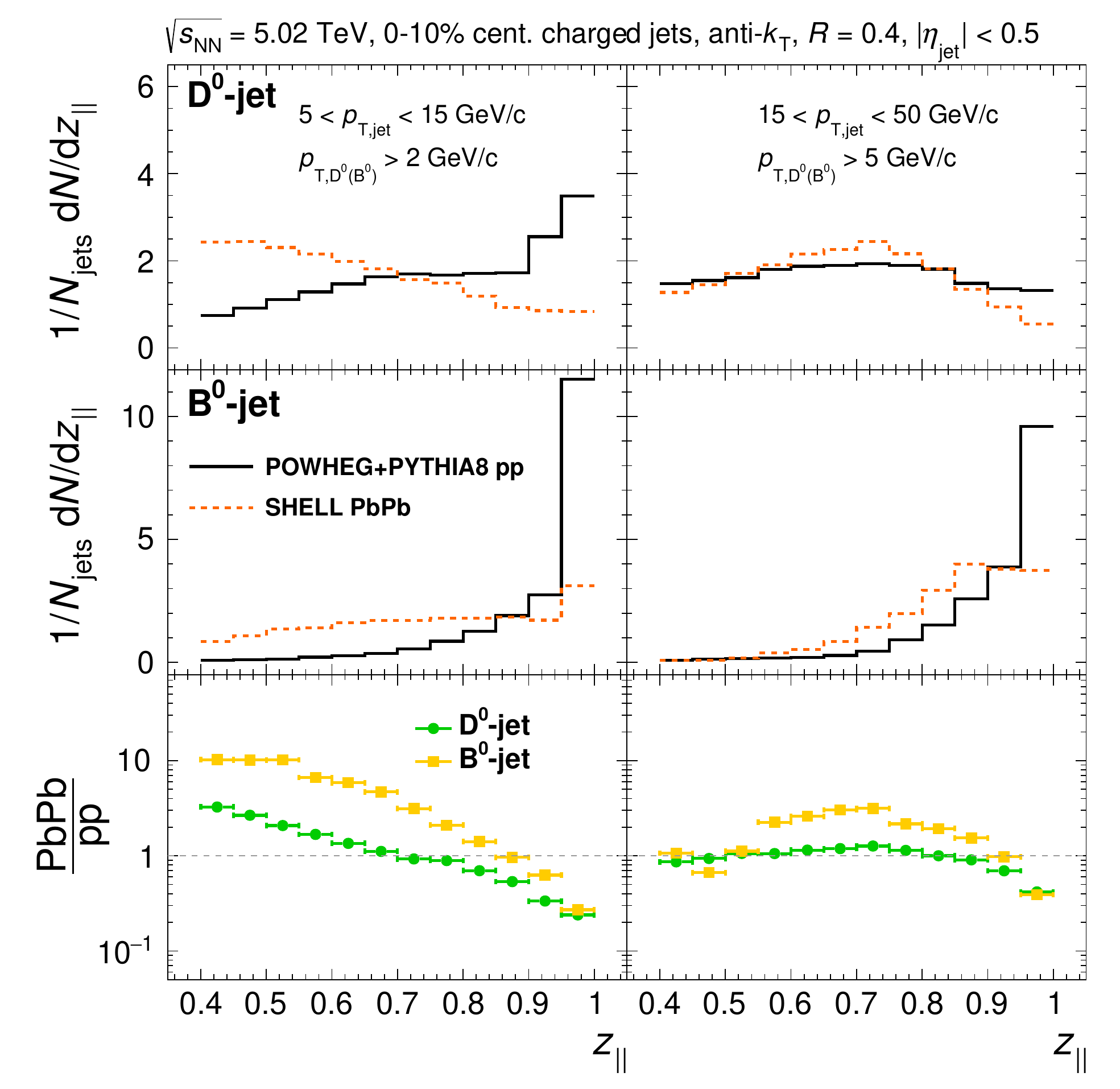}
\vspace*{-.1in}
\caption{$D(z_{||})$ distributions of D-jet and B-jet within two $p_T$ windows both in p+p and 0--10\% Pb+Pb collisions, as well as the medium modifications (PbPb/pp). Figure is from Ref.~\cite{Li:2022tcr}.}
\label{fig:zll}
\end{figure}

On the one hand, the $D(z_{||})$ distribution may provide useful information to reveal the production mechanisms and substructure of heavy quark jets~\cite{ALICE:2022mur}. On the other hand, since $z_{||}$ denotes the momentum projection of the heavy-flavour hadron on the jet axis, the medium modification of the $D(z_{||})$ distribution in nucleus--nucleus collisions is closely related to the interplay of the partonic energy loss between the massive heavy quarks and the massless light partons~\cite{Li:2022tcr}. Figure~\ref{fig:zll} shows the first theoretical investigation of the medium modification on the $D(z_{||})$ distributions of both D-jets and B-jets in Pb+Pb collisions at $\sqrt{s_{NN}}$~=~5.02 TeV. In these calculations, two jet $p_T$ ranges are chosen, \linebreak $5<p_T^{\rm jet}<15$ GeV and $15<p_T^{\rm jet}<50$ GeV. Respectively the selected $D^0$ ($B^0$) mesons are also required to have $p_{T,D^0(B^0)}>$ 2 GeV and $p_{T,D^0(B^0)}>$ 5 GeV. The black solid lines represent the p+p baseline of $D(z_{||})$ distributions calculated by the POWHEG+PYTHIA8 event generator~\cite{Nason:2004rx,Frixione:2007vw,Alioli:2010xd,Sjostrand:2007gs}, and the orange dash lines are the theoretical calculations in Pb+Pb collisions based on the SHELL model. The upper and middle panels correspond to the $D(z_{||})$ distributions of $D$-jets and $B$-jets, while the lower panels are their nuclear modification $D(z_{||})_{PbPb}/D(z_{||})_{pp}$ (green is D-jet and yellow B-jet). One can observe that the initial $D(z_{||})$ distributions in p+p are sensitive to the kinematic region of jet and heavy-flavour hadron, especially for D-jets. Moreover, even within the same kinematic region, a B-jet has an evident harder fragmentation pattern compared to a D-jet. The difference could be relevant to the fact that the stronger ``dead-cone'' effect suffered in heavier bottom quarks, in other words, the bottom quarks radiate less gluon and carry more energy fraction of jets than charm quarks. Besides, the contribution of the GSP process may also play different roles in the production of B-jets and D-jets, which may lead to additional differences in their $z_{||}$ distributions~\cite{Li:2022tcr}. In nuclear collisions, the main finding is that the jet quenching effect results in softer fragmentation patterns of heavy-flavour jets in the QGP compared to that in a vacuum. It's different from what one could naively argue, that is, the energy fraction of heavy quarks in jets may increase because heavy quarks lose less energy than light partons. The modification of $D(z_{||})$ reveals the different energy loss mechanisms between the single parton and the full jet. Critically, the lost energy from the jet constituents may be partially brought back to the jet energy by the reconstruction procedures. This is an essential difference in energy loss mechanisms between the full-jet and the single parton, which leads to less energy loss of full-jet compared to heavy quarks. Therefore, stronger medium modification of $D(z_{||})$ can be obtained with larger R, which may be related to the R-dependence of jet energy loss~\cite{ATLAS:2012tjt,Bossi:2022fpc}. Furthermore, stronger medium modification of $D(z_{||})$ is observed for B-jets compared to D-jets, due to their different initial spectra.
\vspace{-10pt}

\subsection{The ``Dead-Cone'' Effect and Other Observables}

Until now, there are a few other heavy-flavour jet observables accessible in the current experimental measurements at the LHC, which have also attracted attention from the high-energy nuclear physics community. We briefly discuss them in the following.

\begin{itemize}
    \item The Cambridge-Aachen (CA) declustering techniques~\cite{Dokshitzer:1997in} which can help to obtain the angular-ordered pairwise tree of subjets~\cite{Larkoski:2014wba} and the Soft Drop condition mentioned above enable us to expose the most basic heavy quark splitting structure by measuring the splitting-angle distributions in D$^0$ meson jets in p+p collisions at $\sqrt{s_{NN}}=13$~TeV~\cite{ALICE:2021aqk}. It has been measured in three different energy intervals of the radiators: $5 \leq E_{\rm Radiator}  \leq 10$~GeV, $10 \leq E_{\rm Radiator}  \leq 20$~GeV and $20 \leq E_{\rm Radiator}  \leq 30$~GeV and constrain the transverse momentum of the D$^0$ meson in jet to be $2 < p^{\rm D^0}_{\rm T}< 36$~GeV/c. The ALICE collaboration directly observed for the first time a clear distribution suppression at the splitting angle smaller than the ratio of quark mass and the energy of such quark radiator: $\theta \leq M_{\rm charm}/E_{\rm radiator}$, known as the ``dead-cone'' effect~\cite{Dokshitzer:1991fd,Cunqueiro:2018jbh}. Such a heavy quark jet and its substructure measurement reveal and confirm this most basic property of a fast quark interacting with the vacuum described by the QCD theory.

A subsequent phenomenology study exposed the ``dead-cone'' effect of the medium-induced gluon radiation of jet queching~\cite{,Dokshitzer:2001zm,Zhang:2003wk,Armesto:2003jh}, by calculating the emission angle distribution of the heavy-flavour quark initiated splittings in a $\rm D^0$ meson tagged jet and that of the light parton initiated splittings with the existence of the QGP in Pb+Pb collisions at $\sqrt{s_{NN}}=5.02$~TeV ~\cite{Dai:2022sjk}, as demonstrated in Figure~\ref{fig:dcinAA}. Very interestingly, they find the collisional energy loss mechanism will not obscure the observation of the ``dead-cone'' effect in the medium-induced radiation. Such a proposal has also been verified by an analytical study that proposes a new jet substructure groomer that selects the most collinear splitting in a QCD jet above a certain transverse momentum cutoff~\cite{Cunqueiro:2022svx}. It's also found in another study that the ``dead-cone'' domain would be partially filled by the medium-induced emission as heavy quarks traversing QGP~\cite{Armesto:2003jh}.
\begin{figure}[H]
    \hspace{-15pt}    \includegraphics[scale=0.745]{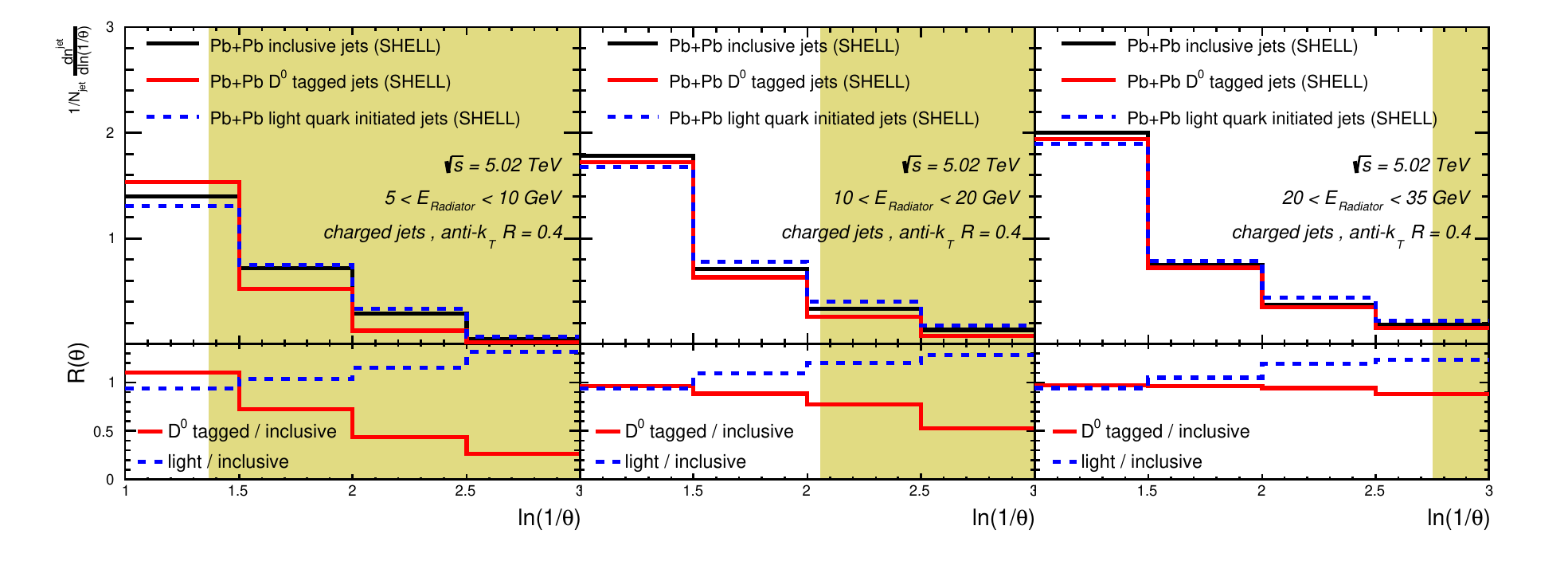}
	\caption{The splitting-angle distributions for $\rm D^0$ meson tagged jets, inclusive jets and also light-quark jets normalized to the number of jets in Pb+Pb collisions at $\rm \sqrt{s}=5.02$~TeV (upper plots) and also the $D^0$ meson tagged jets/inclusive jets (light-quark jets/inclusive jets) ratios (bottom plots) calculated for three energy intervals of the radiators: $\rm 5<E_\text{Radiator}<10$~GeV (left panel),  $\rm 10<E_\text{Radiator}<20$~GeV (middle panel) and  $\rm 20<E_\text{Radiator}<30$~GeV (right panel). The shaded areas correspond to the angles at which the radiation is suppressed due to the ``dead-cone'' effect. Figure is from Ref.~\cite{Dai:2022sjk}.}
	\label{fig:dcinAA}
\end{figure}

 \item The jet shape $\rho(r)$ describes the transverse energy profile of charged hadrons as a function of the angular distance from the jet axis. This observable has been well-studied for light flavor jets~\cite{Chang:2019sae,Luo:2018pto} to search the medium response effect as energetic parton dissipating energy to the medium~\cite{Cao:2020wlm}. The measurement of the medium modification on the b-jet shape has been reposted in Refs.~\cite{CMS:2020geg,CMS:2022btc} by the CMS collaboration. On the ond hand, the comparison of jet shapes of b-jets in Pb+Pb and p+p collisions shows the presence of the QGP modifies the energy distributions around the jet axis of b-jets. On the other hand, their measurements indicate a stronger jet energy redistribution of b-jets at larger radii compared to that of inclusive jets. Generally speaking, the bottom quarks are expected to dissipate less energy in nuclear collisions compared to light quarks and gluons due to the ``dead-cone'' effect. However, at larger jet radii, the medium response effect plays the dominant role in the enhancement of jet energy distribution in Pb+Pb collisions compared to the p+p baseline. Therefore, these interesting results may suggest that the heavier quark, like the bottom,  may drive a stronger medium response effect than a massless parton. In this context, the heavy-flavour jets can serve as promising sensitive probes to the quasi-particle excitation of the quark~soup.

\item The Soft Drop (SD) grooming procedures reveal the two-prong structure of a jet, described by the momentum sharing $z_g$ and opening angle $R_g$~\cite{Larkoski:2014wba}, which establishes the connection between the final state observable to the parton splitting function. The splitting history could be helpful to identify the production mechanisms of heavy-flavour jets~\cite{Goncalves:2015prv,Ilten:2017rbd}. Heavy quark jets from the gluon splitting process usually tend to have more balanced $z_g$ and larger $R_g$ compared to that from the FEX and FCR. The first measurement of the D-jet splitting function is performed by ALICE~\cite{ALICE:2022phr}, and some theoretical efforts which focus on the medium modifications of $z_g$ and $R_g$ of c- and b-jets are presented in Refs. \cite{Li:2017wwc,Zhang:2023jpe}. The medium effects result in more imbalanced $z_g$ distribution and larger opening angles between the two subjets in the heavy quark jets, similar to the medium modification of inclusive jets observed by the CMS~\cite{Sirunyan:2017bsd} and ALICE~\cite{Acharya:2019djg} collaboration.

\end{itemize}

\section{Summary and Conclusions}
\label{sec:summary}

This review covers the current development of theoretical studies on heavy-flavour jets in ultra-relativistic heavy-ion collisions. We introduce the recent theoretical advances of heavy-flavour production in heavy-ion collisions and then give a comprehensive discussion of several recent investigations relating to the heavy-flavour jet observables.

\begin{itemize}

\item We briefly overview the recent theoretical advances that help us understand the heavy-flavour production in heavy-ion collisions, mainly focusing on the initial production,  transport approaches, hadronization mechanism, and diffusion coefficient extraction. These phenomenological studies based on the transport models reveal a fact that the elastic scattering of heavy quarks is dominant at lower $p_T$ region ($p_T^Q<5m_Q$), while the inelastic one dominate the high $p_T$ regions. Besides, different from the fragmentation hadronization of heavy quarks in a vacuum, within the hot and dense nuclear matter, the coalescence mechanism plays an important role in explaining the large collective flow and the enhancement of baryon-to-meson ratio of a charmed hadron in nucleus--nucleus collisions at the RHIC and the LHC. The diffusion coefficient of heavy quarks in the QGP has been extracted by various theoretical frameworks, which implies that $2\pi TD_s$ slightly increases with temperature. The newly developed Bayesian inference approach may be promising to implement a robust determination of the transport coefficient of heavy quarks by a model-data fit.

\item The studies on yield suppression and momentum imbalance of heavy-flavour jets are dedicated to addressing the mass effect of jet energy loss. Theoretical investigations predict stronger yield suppression of light quark jets compared to heavy-flavour jets, which is preliminarily proven by the recent ATLAS measurement of b-jet $R_{AA}$. However, the dijet asymmetry shows a reduced sensitivity to the jet quenching effect, therefore the difference of the medium modification on $x_J$ between inclusive and $b\bar{b}$ dijets seems to be moderate. We have to say the nuclear modification factor is still an effective and powerful observable to test the mass effect of energy loss in QGP. On the other hand, the strategy to isolate the jets initiated by heavy quarks is also crucial to address the mass effect, since GSP processes indeed have a large contribution to the production of heavy quark jets but suffer stronger suppression in nucleus--nucleus~collisions.

\item An observable related to angular correlation aims at the deflection of the jet axis caused by the medium-induced $p_T$-broadening of jet quenching. It's found that the angular deviation caused by the in-medium scattering is hard to be observed for high-$p_T$ jets, both for $b\bar{b}$ dijets and $Z^0\,+\,$(b-)jet. That makes sense because higher $p_T$ jets are more difficult to be changed by the in-medium scattering with the thermal parton in QGP. Meanwhile, medium modification on the radial profiles of jets containing lower-$p_T$ D meson can well capture the angular de-correlation of the charm quark and the jet axis. This suggests that heavy flavors may be more suitable to address the medium-induced $p_T$-broadening of jet quenching since they are experimentally accessible to the low-$p_T$ domain where the angular deviation is visible.

\item The substructure observable can reveal a wealth of information about the inner configuration of heavy-flavour jets. In the vacuum case, declustering techniques provide an inventive way to reestablish the splitting history of hard partons which helps us unlock the ``dead-cone'' effect of charm quark in the experiment. For heavy-flavour jets, the substructure observable also provides a unique opportunity to identify their production mechanisms. Furthermore, jet substructure, such as jet shape, seems more sensitive to the induced medium excitation in nucleus--nucleus collisions than full-jet observables. Much theoretical effort should be made to address the interplay of the ``dead-cone'' effect of medium-induced radiation and the medium response of heavy quarks. From the current perspective, the studies of substructures of heavy-flavour jets could play an increasingly important role in high-energy nuclear physics.

\item The initial jet spectra and the ``selection bias'' play important roles in the medium modifications of jet substructure in nuclear collisions. Normally when we focus on the mass effect of the yield or substructure modification of heavy quark jets, it is apriori to believe that bottom jets should have a weaker medium modification in heavy-ion collisions compared to charm jets under the same conditions. However, in the studies of radial profile and fragmentation function of heavy-flavour jets, it's found that b-jets have very different initial substructure compared to that of c-jets event within the same kinematic constraints, which eventually leads to stronger medium modification of b-jets at the final-state compared to c-jets. On the other hand, the ``selection bias'' poses a challenge to the theoretical studies that aim at the nuclear modification mechanism of heavy-flavour jets in the hot and dense QCD medium. It brings additional ``modifications'' to the ratio PbPb/pp of jet substructure distributions, nevertheless, these ``modifications'' do not exactly reflect the change of jet substructure but only the decrease of jet energy from the higher kinematic region in Pb+Pb collisions.
\end{itemize}




\authorcontributions{{Conceptualization, S.W. and B.-W.Z.; methodology, W.D. and B.-W.Z.; investigation, S.W. and W.D.; writing---original draft preparation, S.W. and W.D.; writing---review and editing, B.-W.Z., E.W. and X.-N.W.; supervision, B.-W.Z., E.W. and X.-N.W.} All authors have read and agreed to the published version of the manuscript.}

\funding{{This research} is supported by the Guangdong Major Project of Basic and Applied Basic Research No.~2020B0301030008, and the Natural Science Foundation of China with Project Nos.~11935007, 12035007, 12247127. S. Wang is also supported by China Postdoctoral Science Foundation under project No.~2021M701279.}

\dataavailability{Not applicable.} 


\conflictsofinterest{The authors declare no conflict of interest.}



\begin{adjustwidth}{-\extralength}{0cm}

\reftitle{References}

\PublishersNote{}
\end{adjustwidth}

\end{document}